\documentclass[preprint]{aastex}

\shorttitle{Primordial Helium and the SMC}
\shortauthors{Peimbert, Peimbert, \& Ruiz}

\begin{document}

\title{The Chemical Composition of the Small Magellanic Cloud H~{\sc{ii}}
Region NGC~346 and the Primordial Helium Abundance}

\author{Manuel Peimbert\altaffilmark{1} and Antonio Peimbert}
\affil{Instituto de Astronom\'{\i}a, Universidad Nacional Aut\'onoma de
M\'exico}

\and

\author{Mar\'{\i}a Teresa Ruiz\altaffilmark{1}}
\affil{Departamento de Astronom\'{\i}a, Universidad de Chile}
    
\altaffiltext{1}{Visiting Astronomer, Cerro Tololo Inter-American Observatory,
which is operated by the Association of Universities for Research in Astronomy,
Inc., under cooperative agreement with the National Science Foundation.}

\begin{abstract}
Spectrophotometry in the $\lambda\lambda$ 3400-7400 range is presented for 13
areas of the brightest H~{\sc{ii}} region in the SMC: NGC 346. The observations
were obtained at CTIO with the 4-m telescope. Based on these observations its
chemical composition is derived. The helium and oxygen abundances by mass are
given by: $Y$(SMC)$ = 0.2405 \pm 0.0018$ and $O$(SMC)$ = 0.00171 \pm 0.00025$.
{From} models and observations of irregular and blue compact galaxies it is
found that $\Delta Y/\Delta O = 3.5 \pm 0.9$ and consequently that the
primordial helium abundance by mass is given by: $Y_p = 0.2345 \pm 0.0026 (1
\sigma)$.  This result is compared with values derived from Big Bang
nucleosynthesis, and with other determinations of $Y_p$.
\end{abstract}


\section{Introduction}

The determination of $Y_p$ based on the Small Magellanic Cloud can have at
least four significant advantages and one disadvantage with respect to those
based on distant H~{\sc{ii}} region complexes: a) no underlying absorption
correction for the helium lines is needed because the ionizing stars can be
excluded from the observing slit, b) the determination of the helium ionization
correction factor can be estimated by observing different lines of sight of a
given H~{\sc{ii}} region, c) the accuracy of the determination can be estimated
by comparing the results derived from different points in a given H~{\sc{ii}}
region, d) the electron temperature is generally smaller than those of metal
poorer H~{\sc{ii}} regions reducing the effect of collisional excitation from
the metastable 2~$^3$S level of He~{\sc{i}}, and e) the disadvantage is that
the correction due to the chemical evolution of the SMC is in general larger
than for the other systems.

The determination of the pregalactic, or primordial, helium abundance by mass
$Y_p$ is paramount for the study of cosmology, the physics of elementary
particles, and the chemical evolution of galaxies \citep[e. g.][and references
therein]{fie98,izo99,pei99}. In this paper we present a new determination of
$Y_p$ based on observations of the SMC. This determination is compared with
those carried out earlier based on extremely metal poor extragalactic
H~{\sc{ii}} regions.

\section{Observations}

Long slit spectra were obtained at CTIO during two observing runs, in August
and September 1990, with the 4-m telescope equipped with the R-C Spectrograph
and a coated GEC CCD detector. Using three different gratings (at first order)
the spectral ranges between $\lambda\lambda$ 3440-5110, 4220-7360, and
5800-7370 were covered. The slit, oriented E-W, was 4.7 arcmin long and 1.6
arcsec wide; the scale along the dispersion axis was 0.73"/pixel.  The
resolution was 7 \AA \ for the blue and red wavelength ranges and 14 \AA \ for
the intermediate one. The slit was placed at two different positions of the
nebula, five extraction windows were defined in one slit position (regions 1-5)
and eight in the other (regions 11-18), four of the extraction windows included
the brightest ionizing stars (m$\sim$14), while the other 9 positions
avoided stars brighter than m = 17 to minimize the stellar contamination of the
nebular spectra. Table~\ref{positions} presents the positions and sizes of the
extraction windows.

In Table~\ref{positions} we also present the observed H$\beta$ fluxes,
$F$(H$\beta$), before correction for extinction. \citet*{mas89} present
excellent pictures of the stellar cluster of NGC 346 where the ionizing stars
included in Table~\ref{positions} are indicated, and the observed regions can
be located. \citet*{ye91} also present excellent pictures of NGC 346 where the
filamentary structure of the nebula can be appreciated. Also in
Table~\ref{positions} we define regions A and B. Region A is the sum of regions
2, 3, 5, 12, 13, 17, and 18. Region A was defined to minimize the errors of the
emission lines and the effect of stellar underlying absorption; the main
results of this paper will be based on it. Regions 1 and 11 were not included
in region A because they are fainter than the others (see
Table~\ref{positions}) and the equivalent widths of their emission lines are
also smaller (see the last paragraph of this section), both effects could
increase the systematic errors. Region B is the sum of all observed regions and
will be used to show the effect of the underlying absorption on the measured
line intensities.

To flux calibrate the spectra five or six spectrophotometric standards, from
\citet{sto83}, were observed each night with the slit widened to 6.4". A
He-Ne-Ar lamp was used to perform the wavelength calibration. Dome-flats and
sky-flats were obtained to flatten the red frames while a quartz lamp flat and
a sky flat were used for the blue range frames. Several bias frames were taken
each night. Data reduction was performed using the IRAF reduction package. Sky
subtraction was made from observations taken one degree away from the nebula
before and after each nebular observation. Short exposure frames at each
position were used to remove cosmic rays and to measure fluxes of strong lines
saturated in the longer exposures.

In Figures \ref{blue}, \ref{red} and \ref{red-detail} we show spectra of region
A at different wavelength ranges. In Figure~\ref{blue-detail} we compare
regions A and B near the Balmer limit to show the effect of the underlying
absorption on the Balmer lines and on $\lambda$ 4026 of He~{\sc{i}}.

In Tables \ref{int-a}, \ref{int-355}, and \ref{int-324} we present the
intrinsic line intensities, $I(\lambda)$, given by
\begin{equation}
\log \left( \frac{I(\lambda)}{I({\rm H}\beta)} \right)=
\log \left( \frac{F(\lambda)}{F({\rm H}\beta)} \right)+
C({\rm H}\beta)f(\lambda),
\end{equation}
where $F(\lambda)$ is the observed line flux corrected for atmospheric
extinction and $C$(H$\beta$) is the logarithmic reddening correction at
H$\beta$, and $f(\lambda)$ is the reddening function. For $f(\lambda)$ we
adopted the normal extinction law \citep{whi58}. $C$(H$\beta$) was obtained by
fitting the observed Balmer decrement with that computed by \citet{bro71} for
$T_e$ = 12 000 K, and $N_e$ = 100 cm$^{-3}$; the Balmer decrement is almost
insensitive to the expected variations in $T_e$ and $N_e$ over the observed
volumes. Table~\ref{int-a} presents the line intensities for regions A and B
that were obtained after adding the spectra of all their components, while
Tables \ref{int-355} and \ref{int-324} present the line intensities for regions
1-5 and 11-18 respectively.

The $C$(H$\beta$) value for region A amounts to $0.15 \pm 0.01$, a value in
good agreement with the values derived from the stellar data of the cluster by
\citet*{nie86} and \citet{mas89} that amount to 0.20 and 0.18 respectively. We
have adopted the $C$(H$\beta$) value of region A for all the observed regions.
The differences between the adopted $f(\lambda)$ and the reddening function in
the visual region in the direction of NGC 346 are small, moreover their effect
on the determination of the line intensity ratios is negligible because with
the adopted $C$(H$\beta$) value we have recovered the theoretical Balmer
decrement of the four brightest lines that are not expected to be affected by
underlying absorption.

The SMC reddening law in the UV is very different to the normal Galactic one,
but fortunately in the visual both laws are similar \citep{bou85}; according to
\citet{bou85} $R = A_V/E(B-V)$ amounts to $2.7 \pm 0.2$ for the SMC. From the
use of various reddening laws with $R$ in the 3.0 to 3.2 range
\citep{whi58,nan75,sea79} we estimate that the error introduced by the adopted
reddening law is about 0.002~dex for all line ratios and smaller than 0.001~dex
for ratios of lines closer than 500 \AA.
 
In Tables \ref{int-a}, \ref{int-355}, and \ref{int-324} one-$\sigma$ errors are
included, the errors were estimated by comparing the results derived from the
two different observing seasons.  The total number of photons received by each
of the He~{\sc{i}} lines of region A is in the $3 \times 10^4$ to $6 \times
10^5$ range, therefore the errors presented in Table~\ref{int-a} are about a
factor of five larger than those given by photon statistics. The errors
presented for region~A are larger because they include other sources of
error present in the reduction procedure.

The line intensities in Tables \ref{int-a}, \ref{int-355}, and \ref{int-324}
were not corrected for underlying stellar absorption with the exception of the
H$\alpha$ lines of regions B, 4, 14, 15, and 16, where we adopted an underlying
absorption of 2 \AA; for these regions we adopted the theoretical
$I$(H$\alpha$)/$I$(H$\beta$) value to normalize all the lines to H$\alpha$,
notice that H$\beta$ has an intensity smaller than unity because it has not
been corrected for underlying absorption.

The $\lambda$~3727 line intensity was corrected for the contribution due to H13
and H14 and to $\lambda$~3724 of [S~{\sc{iii}}], these contributions were
estimated from the other Balmer lines and from the $\lambda$~6312 of
[S~{\sc{iii}}]; $I$(4711) of [Ar~{\sc{iv}}] was obtained after subtracting the
expected contribution of $\lambda$ 4713 of He~{\sc{i}} based on the work by
\citet{smi96}; $I$(3889) of He~{\sc{i}} was obtained after subtracting the
expected contribution of H8 based on the work by \citet{bro71}.

In Table~\ref{e-w} we present the H~{\sc{i}} and He~{\sc{i}} equivalent widths
for all the observed regions. The equivalent widths of regions B, 4, 14, 15,
and 16 are strongly affected by the stellar underlying absorption. No
correction due to underlying absorption has been made for any line in this
table.  After correcting region A for extinction, based on the four brightest
Balmer lines, it is found that the weaker Balmer lines (H9 to H12) are not
affected by stellar underlying absorption (see Figure~\ref{blue-detail}),
therefore the He lines are not expected to be affected by underlying
absorption. {From} similar arguments we expect each of the regions included in
region A (2, 3, 5, 12, 13, 17, and 18) to be unaffected by underlying
absorption.

\section{Temperatures and Densities
\label{T&D}}
 
We derived the temperatures and densities presented in Tables \ref{TandD} and
\ref{TandDA} based on the program of \citet{sha95} for forbidden
lines. $T$(O~{\sc{iii}}), $T$(O~{\sc{ii}}), and $N_e$(S~{\sc{ii}}) values were
derived from the 4363/5007, 3727/7325 and 6716/6731 ratios respectively; to
derive $T$(O~{\sc{ii}}) we considered the contribution by recombination to the
$\lambda\lambda$ 7320,7330 line intensities \citep{liu00}. We will define
another temperature given by
\begin{equation}
T({\rm O~{\scriptstyle{II}}~+~O~{\scriptstyle{III}}}) =
\frac { N({\rm O^+}) T({\rm O~\scriptstyle{II}})+
     N({\rm O^{++}}) T({\rm O~\scriptstyle{III}})}
                      {N({\rm O})},
\end{equation}
where $N$(O$^+$) and $N$(O$^{++}$) are obtained in Section~\ref{ICA} (see Table~\ref{TandDA}).

To derive the root mean square density, $N_e$(rms), we adopted the following
equation
\begin{equation}
N_e^2(rms) =
\left( \frac{3d^2}{r^3} \right)
\left( \frac{I({\rm H}\alpha)}{a({\rm H}\alpha)h\nu({\rm H}\alpha)} \right)
\left[1+ \frac{N({\rm He}^+)}{N({\rm H}^+)} +
2 \frac{N({\rm He}^{++})}{N({\rm H}^+)}\right],
\end{equation}
where $a$(H$\alpha$) is the effective recombination coefficient
\citep[e. g.][]{bro71}, $d$ is the distance to the SMC \citep [64~kpc][and
references therein]{rei99}, and $r$ is the radius of the adopted homogeneous
sphere (150 arcsec). {From} the He$^+$/H$^+$ and He$^{++}$/H$^+$ values for
region A presented in the next section, the $a$(H$\alpha$) value for
$T$~=~11950~K, and the $I$(H$\alpha$) measured by \citet{ken86} we find that
$N_e{\rm (rms)}=14cm^{-3}$ (see Table~\ref{TandDA}). It can be shown that, in
the presence of density fluctuations (which is always the case), $N_e$(rms)
provides us with a lower limit for the local density, $N_e$(local). To derive
the abundance ratios we need to use $N_e$(local) values; in the presence of
density fluctuations $N_e^2$(rms) = $\epsilon \ N_e^2$(local), where $\epsilon$
is the filling factor. From our $N_e$(rms) value and the
$N_e$(He~{\sc{ii}})$_{SC}$ present in Table~\ref{TandDA} (defined in the last
paragraph of this section) it follows that 0.01 is a representative value for
$\epsilon$.

{From} the ratio of the Balmer continuum flux to a Balmer line flux it is
possible to derive the temperature $T_e$(Bac), from the Balmer line
emissivities \citep{sto95,hum87,bro71} and the continuum emissivities for the
He~{\sc{i}} and H~{\sc{i}} continuum \citep{bro70} we find the $T_e$(Bac) value
presented in Table~\ref{TandDA}.

By combining $T$(Bac), $T$(O~{\sc{ii}}), and $T$(O~{\sc{iii}}), together with
$N$(O$^+$) and $N$(O$^{++}$) (see Section~\ref{ICA}), and assuming that
$t^2$(O~{\sc{ii}}) = $t^2$(O~{\sc{iii}}), it is possible to determine the mean
square temperature variation $t^2$(H~{\sc{ii}})~$\equiv$~$t^2$, the average
temperature $T_0$(H~{\sc{ii}})~$\equiv$~$T_0$, and $t^2$(O~{\sc{iii}}) over the
observed volume since \citep{pei67}
\begin{equation}
T_0({\rm X^{i+}})=
\frac{\int T_e N_e N({\rm X^{i+}})dV}
{\int N_e N({\rm X^{i+}})dV},
\end{equation}
\begin{equation}
\label{t^2}
t^2({\rm X^{i+}})=
\frac{\int (T_e - T_0({\rm X^{i+}}))^2 N_e N({\rm X^{i+}})dV}
{T_0({\rm X^{i+}})^2 \int N_e N({\rm X^{i+}})dV},
\end{equation}
\begin{equation}
\label{T0Bac}
T_e(Bac) = T_0(1 - 1.67t^2),
\end{equation}
\begin{equation}
\label{T0Oiii}
T({\rm O~\scriptstyle{III}}) = T_0({\rm O~\scriptstyle{III}})
\left[ 1 + \left( \frac{90800}{T_0({\rm O~\scriptstyle{III}})} - 3\right)
\frac{t^2({\rm O~\scriptstyle{III}})}{2} \right],
\end{equation}
\begin{equation}
\label{T0Oii}
T({\rm O~\scriptstyle{II}}) = T_0({\rm O~\scriptstyle{II}})
\left[ 1 + \left( \frac{97300}{T_0({\rm O~\scriptstyle{II}})} - 3\right)
\frac{t^2({\rm O~\scriptstyle{II}})}{2} \right],
\end{equation}
\begin{equation}
\label{T0O}
T_0 = \frac { N({\rm O^+}) T_0({\rm O~\scriptstyle{II}})+
           N({\rm O^{++}}) T_0({\rm O~\scriptstyle{III}})}
                      {N({\rm O})}.
\end{equation}

In region A $T$(O~{\sc{ii}})~$<$~$T$(O~{\sc{iii}}), therefore
$T_0$(O~{\sc{ii}})~$\neq$~$T_0$(O~{\sc{iii}}), and from equation~(\ref{t^2}) it
follows that $t^2>t^2$(O~{\sc{iii}}). In Table~\ref{TandDA}, we present the
$t^2$(Bac,~O~{\sc{ii}}+O~{\sc{iii}}) value for region A.

In the low density and low optical depth limit it follows
that the emissivities of the helium and hydrogen lines are proportional to
powers of the temperature and consequently that $T_e$(He~{\sc{ii}}) is given by
\citep{pei67}
\begin{eqnarray}
\label{T0He}
\nonumber
T_e({\rm He~\scriptstyle{II}}) \ \ 
= \ \ T_e({\rm He~{\scriptstyle{II}}, H~{\scriptstyle{II}}})
& = & T_0 [1 + (\left< \alpha \right> + \beta - 1)t^2/2] \\
& = & T_0(1 - 1.43t^2),
\end{eqnarray}
where $\left< \alpha \right>$ is the average value of the power of the
temperature for the helium lines and $\beta$ for H$\beta$. The $\alpha$ powers
in the low density limit for the $\lambda\lambda$ 3889, 4026, 4388, 4471, 4922,
5876, 6678, 7065, and 7281 lines are -0.72, -0.98, -1.00, -1.02, -1.04, -1.12,
-1.14, -0.55, and -0.60, respectively \citep{smi96}, and $\beta = -0.89$
\citep[e. g.][]{bro71}. In the low density limit and weighted according to the
observational errors we obtain $\left< \alpha \right> = -0.96$; for
$N_e$(He~{\sc{ii}}) = 143 cm$^{-3}$ we obtain $\left< \alpha \right> =
-0.89$. From $t^2$(Bac,~O~{\sc{ii}}+O~{\sc{iii}}) and
$T_0$(Bac,~O~{\sc{ii}}+O~{\sc{iii}}) derived from
equations~(\ref{T0Bac}~-~\ref{T0O}) together with equation (\ref{T0He}) we
obtain $T_e$(He~{\sc{ii}})$_{Bac}=11890$, which is a representative temperature
for the He~{\sc{i}} lines.

Based on nine emission line ratios in the next section we derive
$N$(He$^+$)/$N$(H$^+$), $T_e$(He~{\sc{ii}}) and $N_e$(He~{\sc{ii}})
self-consistently, hereinafter $N_e$(He~{\sc{ii}})$_{SC}$ and
$T_e$(He~{\sc{ii}})$_{SC}$ (which are presented in Table~\ref{TandDA}). In
Table~\ref{TandDA} we also include
$t^2$(He~{\sc{ii}},~O~{\sc{ii}}+O~{\sc{iii}}) derived from equations
(\ref{T0Oiii}), (\ref{T0Oii}), (\ref{T0O}), and (\ref{T0He}); and
$t^2$(Bac,~He~{\sc{ii}},~O~{\sc{ii}}+O~{\sc{iii}}), an average of the two
$t^2$ determinations; $T_e \left<\right.$(He~{\sc{ii}})$\left.\right>$ which is
the average of $T_e$(He~{\sc{ii}})$_{Bac}$ and $T_e$(He~{\sc{ii}})$_{SC}$; and 
$N_e \left<\right.$(He~{\sc{ii}})$\left.\right>$ which is the density that
corresponds to  $T_e \left<\right.$(He~{\sc{ii}})$\left.\right>$.

{From} this discussion and the values in Table~\ref{TandDA} we conclude that
$t^2 \equiv t^2 ({\rm H~\scriptstyle{II}}) = t^2 ({\rm He~\scriptstyle{II}}) =
0.022 \pm 0.008$ and that to a very good approximation $t^2 ({\rm
O~\scriptstyle{III}}) = t^2 ({\rm H~\scriptstyle{II}})$.

\section{Ionic Chemical Abundances
\label{ICA}}

To determine the abundances from collisionally excited lines many authors adopt
a two temperature scheme, with $T$(O~{\sc{iii}}) for the high degree of
ionization zones and $T$(O~{\sc{ii}}) or $T$(N~{\sc{ii}}) for the low degree of
ionization zones. Therefore to compare with the abundances determined from
collisionally excited lines by other authors we will assume also that that the
temperature within the O~{\sc{ii}} and O~{\sc{iii}} zones is constant, that is
$t^2$(O~{\sc{ii}})=$t^2$(O~{\sc{iii}})=0.000. Two points should be stated here:
(1) under the assumption that $t^2$(O~{\sc{ii}})=$t^2$(O~{\sc{iii}})=0.000 and
since $T$(O~{\sc{ii}})$\neq T$(O~{\sc{iii}}), from equation~(\ref{t^2}), it
follows that $t^2 \neq 0.000$; from the $T$(O~{\sc{iii}}) and $T$(O~{\sc{ii}})
values for region A (see Table~\ref{TandD}) and equations (\ref{t^2}) and
(\ref{T0O}), we find that $t^2=0.0013$; (2) the $t^2=0.0013$ abundances are a
lower limit to the real abundances since the assumption that
$t^2$(O~{\sc{ii}})=$t^2$(O~{\sc{iii}})=0.000 implies constant temperature
within the O~{\sc{ii}} and O~{\sc{iii}} zones, which is not the case (see
section~\ref{T&D}). The total abundances for $t^2$(H~{\sc{ii}})=0.022, our
preferred $t^2$ value, will be discussed in section~\ref{TA}.

Therefore we have determined the ionic abundances of the heavy elements for all
the regions observed using $t^2 = 0.0013$, the abundances are presented in
Table~\ref{RegionIons}. The abundances were computed with the program presented
by \citet{sha95}. To determine the O$^{++}$, Ne$^{++}$, S$^{++}$, Ar$^{++}$,
and Ar$^{3+}$ abundances we used $T$(O~{\sc{iii}}), while for N$^+$, O$^+$, and
S$^+$ we used $T$(O~{\sc{ii}}). For $T$(O~{\sc{iii}}) we adopted the value
derived for each region (see Table~\ref{TandD}). On the other hand, it can be
seen in Table~\ref{TandD} that the $T$(O~{\sc{ii}}) temperatures are
systematically higher for those regions that include bright stars and have
relatively smaller $EW$(H$\beta$) values (regions 4, 14, 15, 16, and B) than
for the regions that have larger $EW$(H$\beta$) values; however, this effect is
not real and has to do with the difficulty of establishing a proper continuum
baseline owing to the presence of underlying Balmer lines in absorption in the
3700--3750 \AA \ region, where the $\lambda$3727 lines originate. This effect,
together with large $T$(O~{\sc{ii}}) errors presented in Table~\ref{TandD}
,which are due to other causes, led us to assume for all regions that
$T$(O~{\sc{ii}})$=0.9036 \, T$(O~{\sc{iii}}), the value determined for
region~A.

To estimate the H$\beta$ emissivity we adopted
\begin{equation}
I ({\rm H}\beta)=
I ({\rm H}\beta\cap{\rm O}^+) + I ({\rm H}\beta\cap{\rm O}^{++}),
\end{equation}
\begin{equation}
\frac{I ({\rm H}\beta\cap{\rm O}^+)}{I({\rm H}\beta\cap{\rm O}^{++})}=
\frac{T({\rm O~\scriptstyle{II}})^{-0.89}}
{T({\rm O~\scriptstyle{III}})^{-0.89}}
\frac{N({\rm O}^+)}{N({\rm O}^{++})},
\end{equation}
where $I$(H$\beta\cap$O$^+$) and $I$(H$\beta\cap$O$^{++}$) are the intensities
of H$\beta$ in the O$^+$ and O$^{++}$ regions respectively. We consider our
procedure adequate since the total O abundances derived for all the regions
show a small dispersion despite their different ionization degree (compare for
example the total O abundances of regions 16 and 18 that are practically the
same).
 
To obtain He$^+$/H$^+$ values we need a set of effective recombination
coefficients for the He and H lines, the contribution due to collisional
excitation to the helium line intensities, and an estimate of the optical depth
effects for the helium lines. The recombination coefficients that we used were
those by \citet{sto95} for H, and \citet{smi96} for He. The collisional
contribution was estimated from \citet{kin95} and \citet*{ben99}. From the
intensity of $\lambda$3614 and the computations of \citet{rob73} it was found
that the He~{\sc{i}} singlet lines were produced under case B. The optical
depth effects in the triplet lines were estimated from the computations by
\citet{rob68}.

To derive the He$^+$/H$^+$ value of region A, in addition to the Balmer lines,
we made use of nine He~{\sc{i}} lines, $\lambda\lambda$ 3889, 4026, 4387, 4471,
4922, 5876, 6678, 7065, and 7281 to determine $N_e$(He~{\sc{ii}}) and
$T_e$(He~{\sc{ii}}) self-consistently.  Each of the 9 He~{\sc{i}}/H~{\sc{i}}
line ratios depends on $T_e$, $N_e$, $N({\rm He}^+)/N({\rm H}^+)$, and the
optical depth of 3889 ($\tau_{3889}$), and each dependence is unique. Therefore
we have a system of 9 equations and 4 unknowns. We decided to obtain the best
value for the 4 unknowns by minimizing $\chi^2$. The $\chi^2$ value is given
by:
\begin{equation}
\label{eq-chi}
\chi^2=\sum_{i=1}^9
\left(
\frac{1-\left<N({\rm He^+})\right>/
N({\rm He^+}(\lambda_i, T_e, N_e, \tau_{3889}))}
{\sigma(\lambda_i)/I(\lambda_i)}
\right)^2,
\end{equation}
where $\sigma(\lambda_i)$ is the absolute error in the measurement that can be
obtained from Table~\ref{int-a}, and $N({\rm He^+}(\lambda_i, T_e, N_e,
\tau_{3889}))/N({\rm H^+})$ is the abundance derived from each line for those
parameters. The best $\tau_{3889}$ value is slightly negative which is
unphysical, moreover from CLOUDY models \citep*{pei00z} it is found that
$\tau_{3889}$ is close to zero, therefore we decided to adopt $\tau_{3889}=0.0$
and to use equation~(\ref{eq-chi}) to derive $T_e$, $N_e$, and $N({\rm
He}^+)/N({\rm H}^+)$ self-consistently.  For a system with 9 independent
determinations and 3 unknowns we have 6 degrees of freedom and we expect the
minimum $\chi^2$ to be in the range $ 1.64 < \chi_{min}^2 < 12.59$ at the 90\%
confidence level. The value of $\chi_{min}^2 = 6.53$ found in Table~\ref{chi}
is in excellent agreement with this range. In Table~\ref{RegionHe} we present
He$^+$/H$^+$ values for different temperatures and densities; the temperatures
were selected to include $T$(O~{\sc{iii}}), $T$(Bac), $T$(He~{\sc{ii}})$_{SC}$,
and two representative temperatures; the densities were selected to include the
minimum $\chi^2$ at each one of the five temperatures. The temperature with the
minimum $\chi^2$ is the self-consistent $T$(He~{\sc{ii}}) and amounts to
$11950\pm 560$~K; this temperature is in excellent agreement with the
temperature derived from the Balmer continuum that amounts to $11800 \pm
500$~K, alternatively $T$(O~{\sc{iii}}) amounts to $13070 \pm 100$~K. Notice
that the $\chi^2$ test requires a higher density for a lower temperature,
increasing the dependence on the temperature of the He$^+$/H$^+$ ratio. As
mentioned above, the values in Table~\ref{chi} correspond to the case where
$\tau_{3889}$ equals zero, for higher values of $\tau_{3889}$ the $\chi^2$
values increase. In Table~\ref{chi} we present He$^+$/H$^+$ values for a set of
temperatures and densities.

{From} Table~\ref{chi} we obtain that $T_e$(He~{\sc{ii}})$_{SC} = 11950$~K and
$N_e$(He{~\sc{ii}})$_{SC} = 143$~cm$^{-3}$, which correspond to He$^+$/H$^+ =
0.0793$.

{From} the surface of $T_e$, $N_e$ and He$^+$/H$^+$ values defined by the
condition $\chi^2 = \chi_{min}^2 +1$ the one-$\sigma$ errors presented in
Table~\ref{TandDA} were computed.

In Table~\ref{RegionHe} we present the He$^+$/H$^+$ values for all the observed
regions without underlying stellar absorption.  We have adopted the
$N_e$(He{~\sc{ii}})$_{SC}$ and $T_e$(He{~\sc{ii}})$_{SC}$ values of region A to
determine the He$^+$/H$^+$ values for all the other regions in
Table~\ref{RegionHe} because their observational errors are higher than for
region A and do not permit to obtain $N_e$(He{~\sc{ii}}) and
$T_e$(He{~\sc{ii}}) self-consistently for each region.

\section{Total Abundances
\label{TA}}

In Table~\ref{TotAbun} we present the total abundances of region A for
$t^2$(O~{\sc{ii}})=$t^2$(O~{\sc{iii}})=0.000 and $t^2 = 0.0013$. The gaseous
abundances were obtained from the following equations \citep{pei69}
\begin{eqnarray}
 \frac{N(\rm O)}{N(\rm H)} & =
             & \frac{N({\rm O^+})+N(\rm O^{++})}{N(\rm H^+)},\\
 \frac{N(\rm N)}{N(\rm H)} & =
             & \left( \frac{N({\rm O^+})+N(\rm O^{++})}{N(\rm O^+)} \right)
               \frac{N(\rm N^+)}{N(\rm H^+)},\\
\frac{N(\rm Ne)}{N(\rm H)} & =
             & \left( \frac{N({\rm O^+})+N(\rm O^{++})}{N(\rm O^{++})} \right)
               \frac{N(\rm Ne^{++})}{N(\rm H^+)}.
\end{eqnarray}

The abundances derived for all the other regions based on these equations agree
within the errors with those derived from region A indicating the reliability
of the equations. To obtain the total O abundances we assumed a correction of
0.04~dex due to the fraction of O tied up in dust grains, this fraction was
estimated from the Si/O values derived for the Orion nebula \citep{est98} and
for H{~\sc{ii}} regions in irregular galaxies \citep{gar95}.

To obtain the total S abundance we adopted the following equation
\begin{equation}
\label{sulphur}
\frac{N(\rm S)}{N(\rm H)} = ICF({\rm S}) \frac{N({\rm S^+}) + N(\rm
S^{++})}{N(\rm H+)};
\end{equation}
from Table~\ref{RegionIons} it can be seen that the higher the O ionization
degree the lower the $[N({\rm S^+}) + N({\rm S^{++}})] / N({\rm H^+})$ ratio
indicating, as expected, the increase of S$^{3+}$ with the O ionization
degree. Therefore to obtain the S abundance we decided to take regions 11 and
12 as representative of NGC 346 since they are expected to have the smallest
$ICF$(S) values. Therefore from the $ICF$(S) values computed by \citet{gar89},
the data for regions 11 and 12 in Table~\ref{RegionIons}, and
equation~(\ref{sulphur}) we derived the value presented in Table~\ref{TotAbun}.

To obtain the total Ar abundance we adopted the following equation
\begin{equation}
\frac{N(\rm Ar)}{N(\rm H)} = ICF({\rm Ar}) \frac{N({\rm Ar^{++}}) + N(\rm
Ar^{3+})}{N(\rm H^+)};
\end{equation}
where $ICF$(Ar) includes the Ar$^+$/H$^+$ contribution and according to
\citet{liu00} can be approximated by
\begin{equation}
ICF({\rm Ar}) =\left( 1- \frac{N(\rm O^+)}{N(\rm O)}\right)^{-1}.
\end{equation}

Based on the $T_e$ values presented in Table~\ref{TandDA} we also include in
Table~\ref{TotAbun} the total abundances for $t^2 = 0.022$, our preferred $t^2$
value, following the procedure outlined by \citet{pei69}. The relevant
equations are:
\begin{eqnarray}
\nonumber
\frac{N({\rm X}^{i+})}{N({\rm H}^+)} & = &
\frac{T({\rm H}\beta \{{\rm X}^{i+}\})^{-0.89}T(\lambda,{\rm X}^{i+})^{-0.5}}
{T(\lambda/\lambda',{\rm X}^{i+})^{-1.39}}\\
\nonumber
&& \times \exp \left(
\frac{\Delta E}{kT(\lambda,{\rm X}^{i+})} -
\frac{\Delta E}{kT(\lambda/\lambda',{\rm X}^{i+})}   \right)\\
&&\times \left[ \frac{N({\rm X}^{i+})}{N({\rm H}^+)}\right]_{\lambda/\lambda'}
\end{eqnarray}
where $\Delta E$ is the energy difference between the ground and the excited
levels, $\lambda$ is the wavelength of the nebular line (5007 \AA \ for
O{~\sc{iii}}), $\lambda '$ is the wavelength of the auroral line (4363 for
O{~\sc{iii}}), $T({\rm H}\beta\{{\rm X}^{i+}\})$ is the $T({\rm H}\beta)$ value
in the region where the X$^{i+}$ ion is present and is given by
\begin{equation}
T({\rm H}\beta\{{\rm X}^{i+}\})=
T_0({\rm X}^{i+}) \left( 1 - 0.945 t^2({\rm X}^{i+})\right);
\end{equation}
$T(\lambda, {\rm X}^{i+})$ is the representative temperature of the nebular
line which can be obtained from
\begin{eqnarray}
\nonumber
T(\lambda, {\rm X}^{i+}) & = & T_0({\rm X}^{i+}) \\
&& \times \left( 1 + \left[
\frac{(\Delta E/kT_0({\rm X}^{i+}))^2 - 3\Delta E/kT_0({\rm X}^{i+}) + 3/4}
{\Delta E/kT_0({\rm X}^{i+}) - 1/2}
\right]\frac{t^2({\rm X}^{i+})}{2}\right);
\end{eqnarray}
and [$N$(X$^{i+}$)/$N$(H$^+$)]$_{\lambda/\lambda'}$ is the abundance derived
from $T(\lambda/\lambda')$ ($T$(O~{\sc{iii}}) for O~{\sc{iii}}).
 
It is possible to derive the abundances
for other $t^2$ values by interpolating or extrapolating the abundances
presented in Table~\ref{TotAbun}.

In Table~\ref{TotAbun} we compare the abundances of NGC 346 with an average of
the H{~\sc{ii}} regions in the SMC derived by \citet{duf84} and those of the
Sun and M17. Notice that the abundances derived by Dufour should be compared
with those for $t^2 = 0.0013$, and that no correction for the fraction of O
tied up in dust grains was included by Dufour. Alternatively the M17 values are
for $t^2=0.037$ and 0.08~dex has been added to the O abundance to correct for
the fraction of this element tied up in dust grains, consequently the M17
values can be compared directly with the NGC~346 values for $t^2=0.022$.

The total He/H value is given by:
\begin{eqnarray}
\nonumber
\frac{N ({\rm He})}{N ({\rm H})} & = & \frac {N({\rm He}^0) + N({\rm He}^+) + 
N({\rm He}^{++})}{N({\rm H}^0) + N({\rm H}^+)},\\
& = & ICF({\rm He}) \frac {N({\rm He}^+) + N({\rm He}^{++})}
{N({\rm H}^+)}.                 
\end{eqnarray}
              
The He$^{++}$/H$^+$ ratio can be obtained directly from the 4686/H$\beta$
intensity ratio. In objects of low degree of ionization the presence of neutral
helium inside the H{~\sc{ii}} region is important and $ICF$(He) becomes larger
than 1. The $ICF$(He) can be estimated by observing a given nebula at different
lines of sight since He$^0$ is expected to be located in the outer regions.
Another way to deal with this problem is to observe H{~\sc{ii}} regions of high
degree of ionization where the He$^0$ amount is expected to be negligible.

\citet[see also \citealt{pag92}]{vil88}  defined a radiation softness parameter
given by
\begin{equation}
\zeta = \frac {N({\rm O}^+)N({\rm S}^{++})}{N({\rm S}^+)N({\rm O}^{++})};
\end{equation}                     
\noindent for large values of $\zeta$ the amount of neutral helium is
significant, while for low values of $\zeta$ it is negligible, where the
critical value is around 8. In Table~\ref{RegionHe} we present the $\zeta$
values for all the observed regions, the $\zeta$ values indicate that the
amount of He$^0$ inside the H$^+$ region is negligible.

On the other hand, for ionization bounded objects of very high degree of
ionization the amount of H$^0$ inside the He$^+$ Str\"omgren sphere becomes
significant and the $ICF$(He) can become smaller than 1. This possibility was
firstly mentioned by \citet{shi74} and studied extensively by \citet{arm99},
\citet*{vie00}, and \citet*{bal00}.

To study this possibility we have estimated the $ICF$(He) from three different
methods. (1) We have divided the regions in Table~\ref{RegionHe} in three
groups: H (regions 5, 17, and 18), I (regions 1, 2, 3, and 13) and L (regions
11 and 12), where H, I, and L stand for high, intermediate and low degree of
ionization regions; within the errors the three groups yield the same He/H
ratio, indicating that the $ICF$(He) is very close to unity. (2) \citet{bal00}
have defined the following cutoff
\begin{equation}
([{\rm O~\scriptstyle{III}}]\lambda5007/{\rm H}\beta)_{\rm cutoff}=
(1.139 \pm 0.306) + (2.5 \pm 0.4) {\rm O/H} \times 10^4.
\end{equation}
For observed values higher than the cutoff $ICF$(He) is very close to unity;
from Table~\ref{TotAbun}, and without considering the fraction of O embedded in
dust grains, we have that $\log N$(O)/$N$(H)$ + 12 = 8.11$ and, consequently, a
cutoff of $4.36 \pm 0.60$. Alternatively the observed value is $5.43 \pm 0.03$.
(3) \citet{bal00} also find that for [O~{\sc{iii}}]$\lambda
5007$/[O~{\sc{i}}]$\lambda 6300 \geq 300$, the $ICF$(He) becomes very close to
unity; from Table~\ref{int-a} it is found that $I(5007)/I(6300) > 600$. (It
should be noted that the [O~{\sc{i}}] lines present in Figure~\ref{red} are
blends of telluric and nebular lines even though we subtracted the sky
contribution. Unfortunatelly the [O~{\sc{i}}] line intensities in the sky
varied in a scale of minutes leaving a telluric remnant present, which becomes
apparent because the centroids of the [O~{\sc{i}}] lines are blue shifted by
about 3 \AA \ from the centroid of the other lines.) From these three methods
we conclude that the amount of H$^0$ inside the He$^+$ Str\"omgren sphere is
negligible and in what follows we will adopt an $ICF$(He) = 1.000.

In Table~\ref{YSMC} we present the helium abundance by mass $Y$(SMC), derived
from region A. The $Y$(SMC) values were derived from Table~\ref{chi}, the
He$^{++}$/H$^+$ value (that amounts to $2.2 \times 10^{-4}$ for the $T_e$ range
present in Table~\ref{YSMC}), and the $Z$ values presented in
Table~\ref{Composition}. The $Y$(SMC) error presented in Table~\ref{Composition}
is based only on the results of the self consistent method; by considering the
$T_e$(Bac) measurement the $Y$(SMC) error diminishes from 0.0018 to 0.0013 (and
$Y$(SMC) diminishes by 0.0001).

In Table~\ref{Composition} we present the helium, heavy elements, and oxygen
abundance by mass of NGC 346. The estimated error for $O$ amounts to 0.06~dex.
The $Z$ value was obtained by assuming that $O$ comprises 54\% of the heavy
elements by mass, this fraction was estimated by \citet{car95} from a group of
ten irregular galaxies that includes the SMC.  The estimated error for the $Z$
value amounts to 0.08~dex.

\section{The Primordial Helium Abundance}

To determine the $Y_p$ value from the SMC it is necessary to estimate the
fraction of helium present in the interstellar medium produced by galactic
chemical evolution. We will assume that
\begin{equation}
\label{DeltaO}
Y_p  =  Y({\rm SMC}) - O({\rm SMC}) \frac{\Delta Y}{\Delta O}.
\end{equation}

In a recent review \citet{pei00y} derive $Y_p$ for the SMC from a similar
discussion but using $Z$ instead of $O$ in equation~(\ref{DeltaO}); since the
error in the $O$ determination is smaller than in the $Z$ determination (see
Table~\ref{Composition}) it is better to use $O$ to determine $Y_p$.

To estimate $\Delta Y/\Delta O$ we will consider three observational
determinations and a few determinations predicted by chemical evolution models.

\citet*{pei92} and \citet{est99} found that $Y = 0.2797 \pm 0.0060$ and $O =
0.0083 \pm 0.0012$ for the Galactic H{~\sc{ii}} region M17, where we have added
0.08~dex to the oxygen gaseous abundance to take into account the fraction of
these elements embedded in dust grains \citep{est98}.  By comparing the $Y$ and
$O$ values of M17 with those of region A we obtain $\Delta Y/\Delta O = 5.45
\pm 1.10$. M17 is the best H{~\sc{ii}} region to determine the helium abundance
because among the brightest galactic H{~\sc{ii}} regions it is the one with the
highest degree of ionization and consequently with the smallest correction for
the presence of He$^0$ \citep[i.e. $ICF$(He) is very close to unity
e.~g.][]{pei92,deh00}. It can be argued that the M17 $\Delta Y/\Delta O$ value
is not representative of irregular galaxies, because the yields are heavy
element dependent and the $O$ value is considerably higher than that of the
SMC.

{From} a group of 10 irregular and blue compact galaxies, that includes the LMC
and the SMC, \citet{car95} found $\Delta Y/\Delta O = 4.48 \pm 1.02$, where
they added 0.2~dex to the O/H abundance ratios derived from the nebular data to
take into account the temperature structure of the H{~\sc{ii}} regions and the
fraction of O embedded in dust; moreover they also estimated that $O$
constitutes 54\% of the $Z$ value. \citet{izo98} from a group of 45 supergiant
H{~\sc{ii}} regions of low metalicity derived a $\Delta Y/\Delta Z = 2.3 \pm
1.0$; we find from their data that $\Delta Y/\Delta Z = 1.46 \pm 0.60$ by
adding 0.2~dex to the O abundances to take into account the temperature
structure of the H{~\sc{ii}} regions and the fraction of O embedded in dust;
furthermore from their data we also find that $\Delta Y/\Delta O = 2.7 \pm 1.2$
by assuming that $O$ constitutes 54\% of the $Z$ value.

Based on their two-infall model for the chemical evolution of the Galaxy
\citet*{chi97} find $\Delta Y/\Delta O = 3.15$ for the solar
vicinity. \citet{car00} computed chemical evolution models for the Galactic
disk, under an inside-out formation scenario, based on different combinations
of seven sets of stellar yields by different authors; the $\Delta Y/\Delta O$
spread predicted by her models is in the 2.9 to 4.6 range for the
Galactocentric distance of M17 (5.9 kpc).

\citet{car95}, based on yields by \citet{mae92}, computed closed box
models adequate for irregular galaxies, like the SMC, and obtained $\Delta
Y/\Delta O = 2.95$. They also computed models with galactic outflows of well
mixed material, that yielded $\Delta Y/\Delta O$ values similar to those of the
closed box models, and models with galactic outflows of O-rich material that
yielded values higher than 2.95. The maximum $\Delta Y/\Delta O$ value that can
be obtained with models of O-rich outflows, without entering into contradiction
with the C/O and $(Z {\rm -C-O)/O}$ observational constraints, amounts to 3.5.

\citet*{car99}, based on yields by \citet*{woo93} and \citet{woo95}, computed
chemical evolution models for irregular galaxies also, like the SMC, and found
very similar values for closed box models with bursting star formation and
constant star formation rates that amounted to $\Delta Y/\Delta O = 4.2$. The
models with O-rich outflows can increase $\Delta Y/\Delta O $, but they
predict higher C/O ratios than observed.

{From} the previous discussion it follows that $\Delta Y/\Delta O = 3.5 \pm
0.9$ is a representative value for models and observations of irregular
galaxies.

The $Y_p$ values in Table~\ref{YpSMC} were computed by adopting $\Delta
Y/\Delta O = 3.5 \pm 0.9$. The differences between Tables \ref{YSMC} and
\ref{YpSMC} depend on $T_e$ because the lower the $T_e$ value the higher the
$O$ value for the SMC. In Figure~\ref{eta} we present our $Y_p$ value as well
as the theoretical $Y_p$ value derived from Big Bang nucleosynthesis
computations by \citet*{cop95} for three neutrino species as a function of
$\eta$, the baryon to photon ratio.

\section{Conclusions}

The $Y_p$ value derived by us is significantly smaller than the value derived
by \citet{izo98} from the $Y$ versus O/H linear regression for a sample of 45
BCGs, and by \citet{izo99} from the average for the two most metal deficient
galaxies known (I~Zw~18 and SBS~0335--052), that amount to $0.2443 \pm 0.0015$
and $0.2452 \pm 0.0015$ respectively (see Figure~\ref{eta}).

The difference could be due to systematic effects in the abundance
determinations.  There are two systematic effects not considered by Izotov and
collaborators that we did take into account, the possible presence of H$^0$
inside the He$^+$ region and the use of a lower temperature than that provided
by the [O{~\sc{iii}}] lines. We consider the first effect to be a minor one and
the second to be a mayor one but both should be estimated for each object. For
further discussion of the first effect see the papers by \citet{vie00} and
\citet{bal00}. The second effect was first mentioned by \citet[see also
\citealt{pei95}]{pei69}.

{From} constant density chemically homogeneous models computed with CLOUDY
\citep{fer96,fer98} we estimate that the maximum temperature that should be
used to determine the helium abundance should be 5\% smaller than
$T_e$(O{~\sc{iii}}). Moreover, if in addition to photoionization there is
additional energy injected to the H{~\sc{ii}} region $T_e$(He{~\sc{ii}}) should
be even smaller.

\citet*{lur99} produced a detailed photoionized model of NGC~2363. For the slit
used by \citet*{izo97} they find an $ICF$(He) = 0.993; moreover they also find
that the $T_e$(O{~\sc{iii}}) predicted by the model is considerably smaller
than observed.  {From} the data of \citet{izo97} for NGC~2363, adopting a
$T_e$(He{~\sc{ii}}) 10\% smaller than $T_e$(O{~\sc{iii}}) and $\Delta Y/\Delta
O = 3.5 \pm 0.9$ we find that $Y_p = 0.234 \pm 0.006$.

Similarly, \citet{sta99} produced a detailed model of I~Zw~18 and find that the
photoionized model predicts a $T_e$(O{~\sc{iii}}) value 15\% smaller than
observed, on the other hand their model predicts an $ICF$(He) = 1.00.  {From}
the observations of $\lambda\lambda$ 5876 and 6678 by \citet{izo99} of I~Zw~18,
and adopting a $T_e$(He{~\sc{ii}}) 10\% smaller than $T_e$(O{~\sc{iii}}) we
obtain $Y_p = 0.237 \pm 0.007$; for a $T_e$(He{~\sc{ii}}) 15\% smaller than
$T_e$(O{~\sc{iii}}) we obtain $Y_p = 0.234 \pm 0.007$, both results in good
agreement with our determination based on the SMC. Further discussion of these
issues is presented elsewhere \citep{pei00z}.

The primordial helium abundance by mass of $0.2345 \pm 0.0026 (1 \sigma)$ ---
based on the SMC --- combined with the Big Bang nucleosynthesis computations by
\citet{cop95} for three light neutrino species implies that, at the
one-$\sigma$ confidence level, $\Omega_b h^2$ is in the 0.0065 to 0.0091 range.
For $h = 0.65$ the $Y_p$ value corresponds to $0.015 < \Omega_b < 0.022$, a
value considerably smaller than that derived from the pregalactic deuterium
abundance, $D_p$, determined by \citet{bur98} that corresponds to $0.041 <
\Omega_b < 0.047 (1 \sigma)$ for $h = 0.65$. Our $\Omega_b$ value is in very
good agreement with the low redshift estimate of the global budget of baryons
by \citet*{fuk98} who find $0.015 < \Omega_b < 0.030 (1 \sigma)$ for $h = 0.65$
and is consistent with their minimum to maximum range for redshift $z = 3$ that
amounts to $0.012 < \Omega_b < 0.070$. The discrepancy between $Y_p$ and $D_p$
needs to be studied further.

In addition to the relevance for cosmology an accurate $Y_p$ value permits to
determine $\Delta Y / \Delta O$, ratio that provides a strong constraint for
the models of chemical evolution of galaxies \citep[see for example][]{car95,car99}.

To increase the accuracy of the $Y_p$ determinations we need observations of
very high quality of as many He{~\sc{i}} lines as possible to derive
$T_e$(He{~\sc{ii}}), $N_e$(He{~\sc{ii}}), $\tau_{3889}$, and
$N$(He$^+$)/$N$(H$^+$) self-consistently. We also need observations with high
spatial resolution to estimate the $ICF$(He) along different lines of sight.

\bigskip

We are grateful to S. Torres-Peimbert for her participation on the observations
presented in this paper. It is a pleasure to acknowledge several fruitful
discussions on this subject with: L. Carigi, V. Luridiana, B. E. J. Pagel,
M. Pe\~na, E. Skillman, G. Steigman, S. Torres-Peimbert, and S. Viegas. We are
also grateful to the referee for some excellent suggestions.  MP received
partial support from CONACyT grant 25451-E, MTR received partial support from a
C\'atedra presidencial and Fondecyt grant 1980659.

\clearpage

\begin{deluxetable}{l r@{.}l lr r@{.}l clll}
\tablecaption{Positions, Sizes, and Observed Fluxes
\label{positions}}
\tablewidth{0pt}
\tablehead{
\colhead{Region} & \multicolumn{3}{c}{Position\tablenotemark{a}}  & \colhead{} & 
\multicolumn{2}{c}{Length\tablenotemark{b}} & \colhead{$F$(H$\beta$)} & \colhead{Star\tablenotemark{c}} & 
\colhead{Spectral}\\
\colhead{} & \multicolumn{2}{c}{$\alpha$} & \colhead{$\delta$} & \colhead{} &
\colhead{} & \colhead{} & \colhead{erg~cm$^{-2}$~s$^{-1}$} & \colhead{} & 
\colhead{Type} }
\startdata
 1 & 102&2$\arcsec$ W & 0.0$\arcsec$ && 34&1$\arcsec$  & 6.23-14\\
 2 &  58&0 W & 0.0  && 22&4 & 1.14-13\\
 3 &  24&9 W & 0.0  && 13&6 & 1.63-13\\
 4 &   2&1 E & 0.0  && 18&6 & 2.37-13 & 355 & O3V((f$^*$)) \\
 5 &  86&4 E & 0.0  && 43&6 & 2.54-13 \\
\\ 
11 & 123&9 W & 5.5 S && 24&8 & 4.33-14\\
12 &  77&6 W & 5.5 S && 32&9 & 9.48-14\\
13 &  42&5 W & 5.5 S && 25&8 & 2.52-13\\
14 &  15&5 W & 5.5 S &&  5&6 & 4.51-14 & 324 & O4V((f)) \\
15 &   7&8 E & 5.5 S &&  9&0 & 1.08-13 & 396 & O7V \\
16 &  22&6 E & 5.5 S &&  9&3 & 1.39-13 & 470+476 & O8-O8.5III+? \\
17 &  66&9 E & 5.5 S && 29&2 & 2.22-13 \\
18 &  92&4 E & 5.5 S && 13&1 & 5.70-14 \\
\\
A\tablenotemark{d}  &   .&.   &       &&  .&. & 11.6-13 \\
B\tablenotemark{e}  &   .&.   &       &&  .&. & 17.9-13 \\
\enddata
\tablenotetext{a}{Relative to star 355: $\alpha = 00^{\rm h}
57^{\rm m}19.82^{\rm s}$; $\delta = -72^\circ 26^\prime 40.4^{\prime\prime}$
(1950).}
\tablenotetext{b}{Slit is $1.6^{\prime\prime}$ wide and is oriented E-W.}
\tablenotetext{c}{Star numbers and spectral types from \citealt{mas89}.}
\tablenotetext{d}{Region A is the sum of Regions 2, 3, 5, 12, 13, 17, 18.}
\tablenotetext{e}{Region B is the sum of Regions 1 to 5 and 11 to 18.}
\end{deluxetable}

\clearpage

\begin{deluxetable}{llcrcc}
\tablecaption{Line Intensities for Regions A and B
\label{int-a}}
\tablewidth{0pt}
\small
\tablehead{
\colhead{$\lambda$} & \colhead{Id.} & \multicolumn{2}{c}{Region A} &
\colhead {Region B\tablenotemark{a}} \\
\cline{3-4} \\
\colhead{} &\colhead{} & \colhead{$\log F(\lambda)/F(H\beta)$} &
\colhead{$\log I(\lambda)/I(H\beta)$}}
\startdata

3614      & He~{\sc{i}}     & $-$2.411 & $-$2.361 $\pm$ 0.040 & ... \\
3634      & He~{\sc{i}}     & $-$2.475 & $-$2.425 $\pm$ 0.040 & ... \\
3726+3729 & [O~{\sc{ii}}]   & $-$0.049 & $-$0.003 $\pm$ 0.005 &   +0.058
$\pm$ 0.005 \\
3750      & H12             & $-$1.546 & $-$1.501 $\pm$ 0.010 & ... \\
3771      & H11             & $-$1.421 & $-$1.376 $\pm$ 0.010 & ... \\
3798      & H10             & $-$1.307 & $-$1.264 $\pm$ 0.010 & ... \\
3835      & H9              & $-$1.176 & $-$1.134 $\pm$ 0.007 & $-$1.677
$\pm$ 0.015 \\
3869      & [Ne~{\sc{iii}}] & $-$0.404 & $-$0.364 $\pm$ 0.003 & $-$0.386
$\pm$ 0.003 \\
3889+3889 & He~{\sc{i}}+H8  & $-$0.740 & $-$0.700 $\pm$ 0.004 & $-$0.915
$\pm$ 0.006 \\
3889      & He~{\sc{i}}     & ...      & $-$1.027 $\pm$ 0.008 & ... \\
\\
3967+3970 & [Ne~{\sc{iii}}]+H7
                            & $-$0.568 & $-$0.533 $\pm$ 0.005 & $-$0.719
$\pm$ 0.006 \\
4026      & He~{\sc{i}}     & $-$1.767 & $-$1.733 $\pm$ 0.015 & ... \\
4069+4076 & [S~{\sc{ii}}]   & $-$2.060 & $-$2.029 $\pm$ 0.020 & ... \\
4102      & H$\delta$       & $-$0.621 & $-$0.591 $\pm$ 0.003 & $-$0.702
$\pm$ 0.005 \\
4340      & H$\gamma$       & $-$0.342 & $-$0.322 $\pm$ 0.003 & $-$0.400
$\pm$ 0.004 \\
4363      & [O~{\sc{iii}}]  & $-$1.139 & $-$1.119 $\pm$ 0.005 & $-$1.175
$\pm$ 0.006 \\
4388      & He~{\sc{i}}     & $-$2.351 & $-$2.332 $\pm$ 0.020 & ... \\
4471      & He~{\sc{i}}     & $-$1.432 & $-$1.416 $\pm$ 0.006 & $-$1.657
$\pm$ 0.008 \\
4658      & [Fe~{\sc{iii}}] & $-$2.688 & $-$2.681 $\pm$ 0.050 & ...\\
4686      & He~{\sc{ii}}    & $-$2.581 & $-$2.574 $\pm$ 0.040 & ... \\
\\
4711+4713 & [Ar~{\sc{iv}}]+He~{\sc{i}}
                            & $-$1.856 & $-$1.851 $\pm$ 0.015 & $-$2.028
$\pm$ 0.020 \\
4711      & [Ar~{\sc{iv}}]  & ...      & $-$2.029 $\pm$ 0.025 & ... \\
4740      & [Ar~{\sc{iv}}]  & $-$2.242 & $-$2.238 $\pm$ 0.025 & $-$2.426
$\pm$ 0.030 \\
4861      & H$\beta$        & ~~~0.000 &    0.000 $\pm$ 0.000 & $-$0.023
$\pm$ 0.003 \\
4922      & He~{\sc{i}}     & $-$1.998 & $-$2.000 $\pm$ 0.008 & $-$2.310
$\pm$ 0.030 \\
4959      & [O~{\sc{iii}}]  & +0.267   &   +0.264 $\pm$ 0.002 &   +0.238
$\pm$ 0.003 \\
5007      & [O~{\sc{iii}}]  & +0.739   &   +0.735 $\pm$ 0.002 &   +0.707
$\pm$ 0.003 \\
5876      & He~{\sc{i}} & $-$0.942     & $-$0.973 $\pm$ 0.005 & $-$1.022
$\pm$ 0.006 \\
6300      & [O~{\sc{i}}]    & $<-$2.016\tablenotemark{b}
                            & $<-$2.060\tablenotemark{b}
                            & $<-$2.083\tablenotemark{b} \\
6312      & [S~{\sc{iii}}]  & $-$1.727 & $-$1.771 $\pm$ 0.010 & $-$1.870
$\pm$ 0.012 \\
\\
6563      & H$\alpha$       & +0.501   &   +0.451 $\pm$ 0.002 &   +0.451
$\pm$ 0.000\tablenotemark{c}\\
6584      & [N~{\sc{ii}}]   & $-$1.371 & $-$1.422 $\pm$ 0.005 & $-$1.365
$\pm$ 0.005 \\
6678      & He~{\sc{i}}     & $-$1.474 & $-$1.528 $\pm$ 0.003 & $-$1.582
$\pm$ 0.005 \\
6716      & [S~{\sc{ii}}]   & $-$1.094 & $-$1.149 $\pm$ 0.003 & $-$1.142
$\pm$ 0.004 \\
6731      & [S~{\sc{ii}}]   & $-$1.237 & $-$1.293 $\pm$ 0.004 & $-$1.274
$\pm$ 0.004 \\
7065      & He~{\sc{i}}     & $-$1.615 & $-$1.675 $\pm$ 0.004 & $-$1.696
$\pm$ 0.005 \\
7136      & [Ar~{\sc{iii}}] & $-$1.065 & $-$1.127 $\pm$ 0.003 & $-$1.129
$\pm$ 0.003 \\
7281      & He~{\sc{i}}     & $-$2.135 & $-$2.199 $\pm$ 0.020 & $-$2.295
$\pm$ 0.025 \\
7320+7330 & [O~{\sc{ii}}]   & $-$1.549 & $-$1.614 $\pm$ 0.005 & $-$1.511
$\pm$ 0.007 \\

\enddata
\tablenotetext{a}{Given in $\log I(\lambda)/I({\rm H}\alpha)+0.451$ (see
text).}
\tablenotetext{b}{Upper limit.} 
\tablenotetext{c}{Corrected for underlying absorption.}
\end{deluxetable}

\clearpage

\begin{deluxetable}{l r@{.}l r@{.}l r@{.}l r@{.}l r@{.}l}
\tabletypesize{\footnotesize}
\tablecaption{Emission Line Intensities\tablenotemark{a}
\label{int-355}}
\tablewidth{0pt}
\tablehead{
\colhead{$\lambda$}   & \multicolumn{2}{c}{1} &
\multicolumn{2}{c}{2} & \multicolumn{2}{c}{3} &
\multicolumn{2}{c}{4\tablenotemark{b}} & \multicolumn{2}{c}{5} }
\startdata
3726+3729 & $+$0&225 $\pm$ 0.008 & $+$0&215 $\pm$ 0.007 & $-$0&246 $\pm$ 0.009
	  & $-$0&042 $\pm$ 0.007 & $-$0&225 $\pm$ 0.007 \\
3835 	  & $-$1&203 $\pm$ 0.040 & $-$1&125 $\pm$ 0.025 & $-$1&127 $\pm$ 0.020	
	  & .&.                  & $-$1&127 $\pm$ 0.015 \\
3869 	  & $-$0&472 $\pm$ 0.012 & $-$0&419 $\pm$ 0.008 & $-$0&403 $\pm$ 0.006
	  & $-$0&394 $\pm$ 0.005 & $-$0&323 $\pm$ 0.005 \\
3889+3889 & $-$0&776 $\pm$ 0.020 & $-$0&702 $\pm$ 0.012 & $-$0&695 $\pm$ 0.009
	  & $-$1&154 $\pm$ 0.015 & $-$0&705 $\pm$ 0.007 \\
3889 	  &    .&.               & $-$1&031 $\pm$ 0.025 & $-$1&017 $\pm$ 0.015
	  &    .&.               & $-$1&038 $\pm$ 0.012 \\
3967+3970 & $-$0&695 $\pm$ 0.020 & $-$0&574 $\pm$ 0.010 & $-$0&567 $\pm$ 0.008
	  & $-$0&886 $\pm$ 0.008 & $-$0&552 $\pm$ 0.006 \\
4026      &    .&.    &    .&.   & $-$1&735 $\pm$ 0.040 &    .&.   &   .&.    \\
4102 	  & $-$0&656 $\pm$ 0.015 & $-$0&590 $\pm$ 0.010 & $-$0&577 $\pm$ 0.008
	  & $-$0&844 $\pm$ 0.008 & $-$0&594 $\pm$ 0.006 \\
4340 	  & $-$0&354 $\pm$ 0.008 & $-$0&319 $\pm$ 0.004 & $-$0&323 $\pm$ 0.003
	  & $-$0&434 $\pm$ 0.004 & $-$0&320 $\pm$ 0.004 \\
4363  	  & $-$1&155 $\pm$ 0.020 & $-$1&152 $\pm$ 0.015 & $-$1&172 $\pm$ 0.012
	  & $-$1&171 $\pm$ 0.010 & $-$1&032 $\pm$ 0.007 \\
4471	  & $-$1&384 $\pm$ 0.030 & $-$1&404 $\pm$ 0.020 & $-$1&409 $\pm$ 0.015
	  & $-$1&469 $\pm$ 0.012 & $-$1&422 $\pm$ 0.015 \\
4686 	  & $<-$2&230\tablenotemark{d} & $<-$2&700\tablenotemark{d}
          & $<-$2&873\tablenotemark{d} & $<-$1&974\tablenotemark{d}
          & $-$2&199 $\pm$ 0.040 \\
4711+4713 &    .&.               & $-$2&275 $\pm$ 0.070 & $-$1&950 $\pm$ 0.035
	  &    .&.               & $-$1&697 $\pm$ 0.020 \\
4711	  &    .&.               &    .&.               & $-$2&188 $\pm$ 0.050
	  &    .&.   & $-$1&814 $\pm$ 0.030 \\
4740 	  &    .&.    &    .&.   & $-$2&498 $\pm$ 0.080 
	  &    .&.   & $-$2&035 $\pm$ 0.035 \\    
4861 	  &    0&000 $\pm$ 0.000 & 0&000 $\pm$ 0.000   &    0&000 $\pm$ 0.000 
	  & $-$0&045 $\pm$ 0.003 &    0&000 $\pm$ 0.000\\
4922 	  & $-$2&019 $\pm$ 0.035 & $-$2&004 $\pm$ 0.025 & $-$1&968 $\pm$ 0.020
	  &    .&.   & $-$2&005 $\pm$ 0.015 \\
4959 	  & $+$0&173 $\pm$ 0.004 & $+$0&209 $\pm$ 0.003 & $+$0&264 $\pm$ 0.002
	  & $+$0&232 $\pm$ 0.003 & $+$0&311 $\pm$ 0.002 \\
5007 	  & $+$0&646 $\pm$ 0.004 & $+$0&679 $\pm$ 0.003 & $+$0&736 $\pm$ 0.002
	  & $+$0&704 $\pm$ 0.003 & $+$0&783 $\pm$ 0.002 \\
5876 	  &    .&.    & $-$0&998 $\pm$ 0.020 & $-$0&962 $\pm$ 0.009
	  & $-$1&029 $\pm$ 0.007 & $-$0&970 $\pm$ 0.007 \\
6312	  & $-$1&736 $\pm$ 0.050 & $-$1&694 $\pm$ 0.030 & $-$1&823 $\pm$ 0.025
	  & $-$1&997 $\pm$ 0.030 & $-$1&970 $\pm$ 0.030 \\
6563	  & $+$0&453 $\pm$ 0.004 & $+$0&453 $\pm$ 0.003 & $+$0&453 $\pm$ 0.002
	  & $+$0&451 $\pm$ 0.000\tablenotemark{c} & $+$0&447 $\pm$ 0.002 \\
6584	  & $-$1&227 $\pm$ 0.015 & $-$1&211 $\pm$ 0.010 & $-$1&705 $\pm$ 0.015
	  & $-$1&485 $\pm$ 0.007 & $-$1&574 $\pm$ 0.008 \\
6678	  & $-$1&537 $\pm$ 0.015 & $-$1&509 $\pm$ 0.012 & $-$1&532 $\pm$ 0.009
	  & $-$1&576 $\pm$ 0.007 & $-$1&537 $\pm$ 0.006 \\
6716 	  & $-$0&961 $\pm$ 0.008 & $-$0&966 $\pm$ 0.006 & $-$1&526 $\pm$ 0.009
	  & $-$1&274 $\pm$ 0.005 & $-$1&177 $\pm$ 0.005 \\
6731 	  & $-$1&108 $\pm$ 0.010 & $-$1&110 $\pm$ 0.007 & $-$1&667 $\pm$ 0.012
	  & $-$1&430 $\pm$ 0.006 & $-$1&319 $\pm$ 0.005 \\
7065 	  & $-$1&645 $\pm$ 0.020 & $-$1&643 $\pm$ 0.015 & $-$1&679 $\pm$ 0.012
	  & $-$1&688 $\pm$ 0.009 & $-$1&673 $\pm$ 0.008 \\
7136	  & $-$1&152 $\pm$ 0.010 & $-$1&090 $\pm$ 0.007 & $-$1&115 $\pm$ 0.005
	  & $-$1&160 $\pm$ 0.004 & $-$1&195 $\pm$ 0.005 \\
7320+7330 & $-$1&410 $\pm$ 0.020 & $-$1&444 $\pm$ 0.015 & $-$1&798 $\pm$ 0.015
	  & $-$1&580 $\pm$ 0.010 & $-$1&873 $\pm$ 0.015 \\
\enddata
\tablenotetext{a}{Given in $\log I(\lambda)/I({\rm H}\beta)$.}
\tablenotetext{b}{Given in $\log I(\lambda)/I({\rm H}\alpha)+0.451$ (see
text).}
\tablenotetext{c}{Corrected for underlying absorption.}
\tablenotetext{d}{One sigma upper limit.}
\end{deluxetable}

\clearpage

\begin{deluxetable}{l r@{.}l r@{.}l r@{.}l r@{.}l r@{.}l r@{.}l r@{.}l r@{.}l}
\rotate
\tabletypesize{\tiny}
\tablewidth{584pt}
\tablecaption{Emission Line Intensities\tablenotemark{a}
\label{int-324}}
\tablehead{
\colhead{$\lambda$}    & \multicolumn{2}{c}{11} &
\multicolumn{2}{c}{12} & \multicolumn{2}{c}{13} &
\multicolumn{2}{c}{14\tablenotemark{b}} &
\multicolumn{2}{c}{15\tablenotemark{b}} &
\multicolumn{2}{c}{16\tablenotemark{b}} &
\multicolumn{2}{c}{17} & \multicolumn{2}{c}{18} }
\startdata
3726+3729 & $+$0&129 $\pm$ 0.010 & $+$0&231 $\pm$ 0.007 & $+$0&167 $\pm$ 0.006 
	  & $-$0&115 $\pm$ 0.020 & $+$0&051 $\pm$ 0.012 & $+$0&328 $\pm$ 0.008 
	  & $-$0&224 $\pm$ 0.008 & $-$0&430 $\pm$ 0.020 \\
3835 	  & $-$1&174 $\pm$ 0.040 & $-$1&155 $\pm$ 0.030 & $-$1&163 $\pm$ 0.015
	  &    .&.   &    .&.   &    .&.
	  & $-$1&155 $\pm$ 0.015 & $-$1&116 $\pm$ 0.040 \\
3869 	  & $-$0&488 $\pm$ 0.012 & $-$0&453 $\pm$ 0.008 & $-$0&422 $\pm$ 0.005
	  & $-$0&354 $\pm$ 0.020 & $-$0&494 $\pm$ 0.012 & $-$0&485 $\pm$ 0.012
	  & $-$0&326 $\pm$ 0.005 & $-$0&319 $\pm$ 0.012 \\
3889+3889 & $-$0&718 $\pm$ 0.015 & $-$0&707 $\pm$ 0.012 & $-$0&704 $\pm$ 0.007
	  &    .&.   &   .&.    &   .&.
	  & $-$0&688 $\pm$ 0.010 & $-$0&698 $\pm$ 0.020  \\
3889 	  & $-$1&069 $\pm$ 0.030 & $-$1&043 $\pm$ 0.020 & $-$1&037 $\pm$ 0.012
	  &    .&.   &   .&.    &   .&.
	  & $-$1&002 $\pm$ 0.012 & $-$1&024 $\pm$ 0.035  \\
3967+3970 & $-$0&609 $\pm$ 0.015 & $-$0&555 $\pm$ 0.010 & $-$0&573 $\pm$ 0.006
	  &    .&.   &   .&.    &   .&.
	  & $-$0&539 $\pm$ 0.006 & $-$0&537 $\pm$ 0.015 \\
4069+4076 &    .&.    & $-$1&930 $\pm$ 0.080 & $-$1&988 $\pm$ 0.040 
	  &    .&.   &   .&.    &   .&.
	  & $-$2&131 $\pm$ 0.060 &    .&. \\
4102 	  & $-$0&597 $\pm$ 0.015 & $-$0&593 $\pm$ 0.012 & $-$0&595 $\pm$ 0.006
	  &    .&.   & $-$0&982 $\pm$ 0.025 & $-$0&953 $\pm$ 0.025
	  & $-$0&589 $\pm$ 0.007 & $-$0&583 $\pm$ 0.015 \\
4340 	  & $-$0&323 $\pm$ 0.008 & $-$0&332 $\pm$ 0.005 & $-$0&329 $\pm$ 0.003
	  & $-$0&509 $\pm$ 0.015 & $-$0&496 $\pm$ 0.009 & $-$0&405 $\pm$ 0.007
	  & $-$0&328 $\pm$ 0.003 & $-$0&325 $\pm$ 0.006 \\
4363  	  & $-$1&259 $\pm$ 0.025 & $-$1&264 $\pm$ 0.015 & $-$1&237 $\pm$ 0.010
	  & $-$1&141 $\pm$ 0.040 & $-$1&250 $\pm$ 0.025 & $-$1&272 $\pm$ 0.025
	  & $-$1&085 $\pm$ 0.010 & $-$1&022 $\pm$ 0.020 \\
4471	  &    .&.    & $-$1&410 $\pm$ 0.020 & $-$1&419 $\pm$ 0.012
	  &    .&.   &   .&.    &   .&.
	  & $-$1&415 $\pm$ 0.015  & $-$1&428 $\pm$ 0.035 \\
4686 	  & $<-$2&181\tablenotemark{d}  & $<-$2&662\tablenotemark{d}  
          & $<-$2&815\tablenotemark{d} 
	  &  .&.   &   .&.    &   .&.  & $<-$2&433\tablenotemark{d}
          & $<-$2&609\tablenotemark{d} \\
4711+4713 &    .&.    &    .&.   & $-$2&152 $\pm$ 0.035 &    .&.   &   .&.    &   .&.
	  & $-$1&847 $\pm$ 0.025  & $-$1&694 $\pm$ 0.050 \\
4711	  &    .&.    &    .&.   & $-$2&637 $\pm$ 0.050 &    .&.   &   .&.    &   .&.
	  & $-$2&023 $\pm$ 0.035  & $-$1&810 $\pm$ 0.060 \\
4740 	  &    .&.    &    .&.   &    .&.   &    .&.    &   .&.   &   .&.
	  & $-$2&239 $\pm$ 0.040  & $-$1&965 $\pm$ 0.070 \\    
4861 	  &    0&000 $\pm$ 0.000  &    0&000 $\pm$ 0.000 &    0&000 $\pm$ 0.000
	  & $-$0&077 $\pm$ 0.008  & $-$0&071 $\pm$ 0.005 & $-$0&025 $\pm$ 0.004
	  &    0&000 $\pm$ 0.000  &    0&000 $\pm$ 0.000 \\
4922 	  &    .&.    & $-$1&995 $\pm$ 0.025 & $-$1&982 $\pm$ 0.012 
	  &    .&.    &   .&.   &   .&.
	  & $-$2&023 $\pm$ 0.020  &    .&.  \\
4959 	  & $+$0&205 $\pm$ 0.004 & $+$0&188 $\pm$ 0.003 & $+$0&210 $\pm$ 0.002
	  & $+$0&282 $\pm$ 0.008 & $+$0&151 $\pm$ 0.005 & $+$0&135 $\pm$ 0.004
	  & $+$0&294 $\pm$ 0.002 & $+$0&332 $\pm$ 0.004 \\
5007 	  & $+$0&672 $\pm$ 0.004 & $+$0&660 $\pm$ 0.003 & $+$0&683 $\pm$ 0.002
	  & $+$0&757 $\pm$ 0.008 & $+$0&628 $\pm$ 0.005 & $+$0&597 $\pm$ 0.004
	  & $+$0&767 $\pm$ 0.002 & $+$0&804 $\pm$ 0.004 \\
5876 	  & $-$0&977 $\pm$ 0.020 & $-$0&988 $\pm$ 0.012 & $-$0&987 $\pm$ 0.008
	  & $-$1&074 $\pm$ 0.040 & $-$1&047 $\pm$ 0.020 & $-$1&205 $\pm$ 0.020
	  & $-$0&983 $\pm$ 0.008 & $-$0&981 $\pm$ 0.015 \\
6312	  & $-$1&623 $\pm$ 0.040 & $-$1&672 $\pm$ 0.035 & $-$1&761 $\pm$ 0.015
	  & $-$1&953 $\pm$ 0.100 & $-$1&931 $\pm$ 0.060 & $-$1&961 $\pm$ 0.070
	  & $-$1&869 $\pm$ 0.025 & $-$1&980 $\pm$ 0.040 \\
6563	  & $+$0&445 $\pm$ 0.004 & $+$0&446 $\pm$ 0.003 & $+$0&446 $\pm$ 0.002
	  & $+$0&451 $\pm$ 0.000\tablenotemark{c} & $+$0&451 $\pm$ 
            0.000\tablenotemark{c} & $+$0&451 $\pm$ 0.000\tablenotemark{c}
	  & $+$0&452 $\pm$ 0.002 & $+$0&443 $\pm$ 0.004 \\
6584	  & $-$1&381 $\pm$ 0.020 & $-$1&228 $\pm$ 0.012 & $-$1&273 $\pm$ 0.008
	  & $-$1&546 $\pm$ 0.040 & $-$1&379 $\pm$ 0.020 & $-$1&137 $\pm$ 0.012
	  & $-$1&732 $\pm$ 0.015 & $-$1&769 $\pm$ 0.025 \\
6678	  & $-$1&546 $\pm$ 0.020 & $-$1&542 $\pm$ 0.012 & $-$1&532 $\pm$ 0.007
	  & $-$1&672 $\pm$ 0.040 & $-$1&600 $\pm$ 0.020 & $-$1&719 $\pm$ 0.020
	  & $-$1&526 $\pm$ 0.008 & $-$1&546 $\pm$ 0.020 \\
6716 	  & $-$1&118 $\pm$ 0.012 & $-$0&966 $\pm$ 0.006 & $-$1&041 $\pm$ 0.004
	  & $-$1&396 $\pm$ 0.030 & $-$1&140 $\pm$ 0.012 & $-$0&985 $\pm$ 0.008
	  & $-$1&246 $\pm$ 0.005 & $-$1&512 $\pm$ 0.020 \\
6731 	  & $-$1&306 $\pm$ 0.015 & $-$1&124 $\pm$ 0.007 & $-$1&184 $\pm$ 0.005
	  & $-$1&548 $\pm$ 0.035 & $-$1&251 $\pm$ 0.015 & $-$1&075 $\pm$ 0.009
	  & $-$1&389 $\pm$ 0.006 & $-$1&677 $\pm$ 0.020 \\
7065 	  & $-$1&627 $\pm$ 0.025 & $-$1&690 $\pm$ 0.015 & $-$1&687 $\pm$ 0.010
	  & $-$1&815 $\pm$ 0.050 & $-$1&759 $\pm$ 0.025 & $-$1&907 $\pm$ 0.030
	  & $-$1&669 $\pm$ 0.010 & $-$1&679 $\pm$ 0.020 \\
7136	  & $-$1&097 $\pm$ 0.012 & $-$1&098 $\pm$ 0.007 & $-$1&096 $\pm$ 0.004
	  & $-$1&168 $\pm$ 0.020 & $-$1&174 $\pm$ 0.012 & $-$1&175 $\pm$ 0.010
	  & $-$1&145 $\pm$ 0.005 & $-$1&202 $\pm$ 0.012 \\
7320+7330 & $-$1&585 $\pm$ 0.025 & $-$1&385 $\pm$ 0.015 & $-$1&406 $\pm$ 0.012
	  & $-$1&645 $\pm$ 0.040 & $-$1&486 $\pm$ 0.020 & $-$1&189 $\pm$ 0.020
	  & $-$1&864 $\pm$ 0.015 & $-$1&945 $\pm$ 0.030 \\
\enddata
\tablenotetext{a}{Given in $\log I(\lambda)/I({\rm H}\beta)$.}
\tablenotetext{b}{Given in $\log I(\lambda)/I({\rm H}\alpha)+0.451$ (see
text).}
\tablenotetext{c}{Corrected for underlying absorption.}
\tablenotetext{d}{One sigma upper limit.}
\end{deluxetable}

\clearpage

\begin{deluxetable}{lrrrrrrrrrr}
\tablecaption{Emission Equivalent Widths\tablenotemark{a}
\label{e-w}}
\tablewidth{0pt}
\tablehead{
\colhead{Region}    & \colhead{H$\alpha$}  & \colhead{H$\beta$} &
\colhead{H$\gamma$} & \colhead {H$\delta$} & \colhead{H9} &
\colhead{4471}      & \colhead{4922}       & \colhead{5876} &
\colhead{6678}      & \colhead{7065} }
\startdata
 1 &  520 &  85 &  25 &  11 & 2.2 & 2.5 & 0.9 & 18 & 7.0 & 6.5 \\
 2 &  720 & 190 &  75 &  38 & 10  & 5.5 & 2.0 & 28 & 9.5 & 8.5 \\
 3 & 1650 & 380 & 140 &  65 & 17  & 12  & 4.5 & 60 & 20  & 16 \\
 4 &  115 &  19 & 5.0 & 1.5 & -0.65\tablenotemark{b} & 0.55 & 0.35 & 2.8& 1.2 
& 1.1 \\ 
 5 &  785 & 210 & 90 & 45 & 11 & 7.5 & 2.2 & 28& 8.5 & 7.0 \\
\\
11 &  580 & 140 &  60 & 30  & 7.5 & 5.0 & ... & 22  &  7.0 &  5.5 \\
12 &  790 & 200 &  75 & 38  & 9.5 & 7.2 & 1.4 & 33  &  8.5 &  7.0 \\
13 & 1760 & 350 & 115 & 60  & 13  & 11  & 4.0 & 65  &   21 & 16 \\
14 &   55 & 7.0 & 2.0 & ... & ... & ... & ... & 1.4 & 0.50 &  0.45 \\
15 &  200 &  22 & 4.8 & ... & ... & ... & ... & 4.5 &  2.0 &  1.9 \\
16 &   69 & 8.0 & 2.5 & ... & ... & -0.55\tablenotemark{b} &  ... &
1.0 & 0.29& 0.16 \\
17 & 1950 & 340 & 120 & 55  & 14 & 10   & 3.5 & 70 & 21 & 19 \\
18 & 1800 & 300 & 100 & 60  & 13 & 9.5  & 4.0 & 65 & 20 & 18 \\
\\
A  & 1250 & 250 & 99 & 44 & 12.5 & 8.5 & 2.5 & 46 & 14.7 & 11.2 \\
B  &  230 &  35 & 11.1 & 4.7 & 0.43 & 0.7 & 0.18 & 6.5 & 2.7 & 2.2 \\  
\enddata
\tablenotetext{a}{In Angstroms.}
\tablenotetext{b}{$EW$ in absorption.}
\end{deluxetable}

\clearpage

\begin{deluxetable}{cr@{\ $\pm$\ }rr@{\ $\pm$\ }rr@{\ $\pm$\ }r}
\tablecaption{Temperatures and Densities
\label{TandD}}
\tablewidth{0pt}
\tablehead{
\colhead{Region}&
\multicolumn{2}{c}{$T$(O~{\sc{iii}})}&
\multicolumn{2}{c}{$T$(O~{\sc{ii}})}&
\multicolumn{2}{c}{$N_e$(S~{\sc{ii}})}}
\startdata
1 & 13 730 & 250 & 11 650 &  450 &  40 &  35 \\
2 & 13 350 & 190 & 11 150 &  400 &  50 &  25 \\
3 & 12 500 & 130 & 12 600 &  350 &  55 &  45 \\
4 & 12 840 & 130 & 13 600 &  350 &  15 &  20 \\
5 & 13 540 & 100 & 10 950 &  250 &  55 &  20 \\
\\
11& 12 230 & 250 & 10 500 &  600 &  14\tablenotemark{a} &  50 \\
12& 12 320 & 170 & 12 100 &  350 &  15 &  25 \\
13& 12 360 & 130 & 12 600 &  300 &  50 &  20 \\
14& 12 580 & 390 & 13 550 & 1150 &  25 & 150 \\
15& 12 800 & 270 & 12 550 &  550 & 150 &  60 \\
16& 12 890 & 310 & 12 550 &  600 & 220 &  45 \\
17& 13 120 & 140 & 11 100 &  300 &  55 &  20 \\
18& 13 440 & 280 & 13 500 &  950 &  14\tablenotemark{a} &  65 \\
\\
A & 13 070 &  50 & 11 810 &  160 &  50 &  15 \\
B & 12 755 &  90 & 12 360 &  220 &  85 &  20 \\
\enddata
\tablenotetext{a}{Low density limit.}
\end{deluxetable}

\clearpage

\begin{deluxetable}{cr@{=}lcr@{=}lcr@{=}rc}
\tabletypesize{\small}
\tablecaption{Temperatures and Densities for Region A
\label{TandDA}}
\tablewidth{0pt}
\tablehead{
&\multicolumn{2}{c}{$T_e$($^o$K)} &&
\multicolumn{2}{c}{$t^2$} &&
\multicolumn{2}{c}{$N_e$(cm$^{-3}$)}&}
\startdata
&(O~{\sc{iii}})                                 & $13070\pm100$ &&
	\multicolumn{2}{c}{ } &&
		(rms)                                         & $ 14\pm~3$ \\
&(O~{\sc{ii}})                                  & $11810\pm300$ &&
	\multicolumn{2}{c}{ } &&
		(S~{\sc{ii}})                                 & $ 50\pm15$ \\
&(O~{\sc{ii}}+O~{\sc{ii}i})                     & $12860\pm100$ &&
	\multicolumn{2}{c}{ } &&
		\multicolumn{2}{c}{ } \\
&(Bac)                                          & $11800\pm500$ &&
	(Bac, O~{\sc{ii}}+O~{\sc{iii}})               & $0.023\pm0.011$ &&
		\multicolumn{2}{c}{ } \\
&(He~{\sc{ii}})$_{Bac}$                         & $11890\pm500$ &&
	\multicolumn{2}{c}{ } &&
		\multicolumn{2}{c}{ } \\
&(He~{\sc{ii}})$_{SC}$                          & $11950\pm560$ &&
	(He~{\sc{ii}}, O~{\sc{ii}}+O~{\sc{iii}})      & $0.021\pm0.012$ &&
		(He~{\sc{ii}})$_{SC}$                         & $143\pm50$ \\
&$\left<\right.$(He~{\sc{ii}})$\left. \right>$  & $11920\pm370$ &&
	(Bac, He~{\sc{ii}}, O~{\sc{ii}}+O~{\sc{iii}}) & $0.022\pm0.008$ &&
		$\left<\right.$(He~{\sc{ii}})$\left. \right>$ & $146\pm50$ \\
\enddata
\end{deluxetable}

\clearpage

\begin{deluxetable}{ccccccccc}
\tablecaption{Ionic Abundances\tablenotemark{a} \
for $t^2=0.0013$\tablenotemark{b}
\label{RegionIons}}
\tablewidth{0pt}
\tablehead{
\colhead{Region}    & \colhead{O$^+$}     & \colhead{O$^{++}$}  &
\colhead{N$^+$}     & \colhead{Ne$^{++}$} & \colhead{S$^+$}     &
\colhead{S$^{++}$}  & \colhead{Ar$^{++}$} & \colhead{Ar$^{3+}$} }
\startdata
~1 & 7.44 & 7.79 & 5.87 & 6.95 & 5.42 & 6.11 & 5.52 & .... \\
~2 & 7.47 & 7.86 & 5.91 & 7.04 & 5.44 & 6.19 & 5.61 & .... \\
~3 & 7.10 & 7.99 & 5.45 & 7.15 & 4.93 & 6.16 & 5.64 & 4.65 \\
~4 & 7.27 & 7.93 & 5.67 & 7.12 & 5.15 & 5.95 & 5.57 & .... \\
~5 & 7.01 & 7.94 & 5.53 & 7.11 & 5.21 & 5.89 & 5.49 & 4.98 \\
\\
11 & 7.51 & 7.96 & 5.81 & 7.10 & 5.35 & 6.40 & 5.68 & .... \\
12 & 7.60 & 7.94 & 5.96 & 7.12 & 5.50 & 6.34 & 5.67 & .... \\
13 & 7.53 & 7.95 & 5.91 & 7.15 & 5.42 & 6.25 & 5.66 & 4.56 \\
14 & 7.22 & 8.01 & 5.62 & 7.19 & 5.05 & 6.02 & 5.58 & .... \\
15 & 7.38 & 7.86 & 5.77 & 7.03 & 5.32 & 6.02 & 5.56 & .... \\
16 & 7.65 & 7.82 & 6.01 & 7.03 & 5.49 & 5.98 & 5.55 & .... \\
17 & 7.06 & 7.97 & 5.40 & 7.16 & 5.17 & 6.05 & 5.56 & 4.81 \\
18 & 6.83 & 7.97 & 5.34 & 7.13 & 4.88 & 5.90 & 5.49 & 5.07 \\
\\
~A & 7.28 & 7.94 & 5.69 & 7.13 & 5.27 & 6.15 & 5.59 & 4.81 \\
~B & 7.41 & 7.97 & 5.81 & 7.17 & 5.33 & 6.11 & 5.63 & 4.64 \\
\enddata
\tablenotetext{a}{Given by $\log N({\rm X})/N({\rm H}^+) +12$.}
\tablenotetext{b}{We have adopted a two temperature model with $T$(O~{\sc{ii}})
and $T$(O~{\sc{iii}}), inside the O$^+$ and O$^{++}$ zones the temperature is
uniform ($t^2=0.000$), but the $t^2$ over the entire model turns out to be
0.0013. Notice that our preferred $t^2$ value is 0.022, see Tables \ref{TandDA}
and \ref{TotAbun}.}
\end{deluxetable}

\clearpage

\begin{deluxetable}{ccccccc}
\tablecaption{$N({\rm He}^+)/N({\rm H}^+)$\tablenotemark{a} \
and $\chi^2$ for Region A
\label{chi}}
\tablewidth{0pt}
\tablehead{\colhead{$T_e$(K)} & \ & \multicolumn{5}{c}{$N_e$(cm$^{-3}$)} \\
&& \colhead{53}  & \colhead{100} & \colhead{143}
 & \colhead{162} & \colhead{247}}
\startdata
11200 && 805    & 798    & 793    & 791    & 781                    \\
      && (83.2) & (47.7) & (26.4) & (20.0) & (8.24)\tablenotemark{b}   \\
\\
11800 && 806    & 799    & 793    & 790                       & 780 \\
      && (38.6) & (15.9) & (7.37) & (6.59)\tablenotemark{b}   & (20.4) \\
\\
11950 && 806    & 799    & 793                       & 790    & 779 \\
      && (30.8) & (11.7) & (6.53)\tablenotemark{b,c} & (7.25) & (27.7) \\
\\
12400 && 807    & 799                       & 793    & 790    & 778 \\
      && (15.0) & (7.17)\tablenotemark{b}   & (12.5) & (17.9) & (58.6) \\
\\
13000 && 809                       & 800    & 793    & 790    &  777 \\
      && (9.72)\tablenotemark{b}   & (18.2) & (38.4) & (50.2) & (118) \\
\enddata
\tablenotetext{a}{Given in units of $10^{-4}$, $\chi^2$ values in parenthesis.}
\tablenotetext{b}{The minimum $\chi^2$ value at a given temperature.}
\tablenotetext{c}{The smallest $\chi^2$ value for all temperatures and
densities, thus defining $T_e$(He~{\sc{ii}}) and $N_e$(He~{\sc{ii}}).}
\end{deluxetable}

\clearpage

\begin{deluxetable}{ccccccccl}
\tablecaption{Ionic Helium Abundances\tablenotemark{a}
\label{RegionHe}}
\tablewidth{0pt}
\tablehead{
\colhead{Region}    & \colhead{3889}      & \colhead{4471}      &
\colhead{4922}      & \colhead{5876}      & \colhead{6678}      &
\colhead{7065}      & \colhead{$\left< {\rm He^+/H^+} \right>$} &
\colhead{$\zeta$}    }
\startdata
~1 & ... & 838 & 739 & ... & 778 & 849 & 802 & 2.07  \\
~2 & 802 & 800 & 764 & 755 & 830 & 853 & 815 & 2.21  \\
~3 & 828 & 791 & 831 & 820 & 787 & 784 & 804 & 2.08  \\
~5 & 789 & 768 & 763 & 806 & 778 & 796 & 788 & 0.525 \\
11 & 735 & ... & ... & 793 & 762 & 885 & 794 & 4.03  \\
12 & 780 & 789 & 780 & 763 & 769 & 765 & 773 & 3.10  \\
13 & 791 & 773 & 804 & 775 & 787 & 771 & 783 & 2.36  \\
17 & 850 & 783 & 735 & 786 & 805 & 788 & 797 & 0.873 \\
18 & 815 & 757 & ... & 785 & 763 & 785 & 780 & 0.719 \\
\\
~A & 809 & 779 & 772 & 800 & 795 & 792 & 793\tablenotemark{b} & 1.57 \\
\enddata
\tablenotetext{a}{Given by $10^4 \times N({\rm He}^+)/N({\rm H}^+)$.}
\tablenotetext{b}{Average of 9 lines, see Table \ref{chi}.}
\end{deluxetable}

\clearpage

\begin{deluxetable}{cccccc}
\tablecaption{Total Abundances\tablenotemark{a}
\label{TotAbun}}
\tablewidth{0pt}
\tablehead{\colhead{Element} &
    \colhead{SMC\tablenotemark{b}} &
        \multicolumn{2}{c}{NGC~346\tablenotemark{c}} &
            \colhead{Sun\tablenotemark{d}} &
                \colhead{M17\tablenotemark{e}} \\
&& \colhead{$t^2=0.0013$\tablenotemark{f}}
& \colhead{$t^2=0.0220$\tablenotemark{g}}}
\startdata
N  & 6.46 & 6.44 & 6.51 & 7.92 & 7.90 \\
O  & 8.02 & 8.07 & 8.15 & 8.83 & 8.87 \\
Ne & 7.22 & 7.22 & 7.30 & 8.08 & 8.02 \\
S  & 6.49 & 6.50 & 6.59 & 7.33 & 7.31 \\
Ar & 5.78 & 5.74 & 5.82 & 6.40 & 6.60 \\
\enddata
\tablenotetext{a}{Given by $\log N({\rm X})/N({\rm H}) + 12$.}
\tablenotetext{b}{\citealt{duf84}.}
\tablenotetext{c}{This paper.}
\tablenotetext{d}{\citealt{gre98}.}
\tablenotetext{e}{\citealt{pei92,est99}.}
\tablenotetext{f}{Minimum $t^2$, see first paragraph in section~\ref{ICA}.}
\tablenotetext{g}{Preferred $t^2$, see section~\ref{T&D} and
Table~\ref{TandDA}.} 
\end{deluxetable}

\clearpage

\begin{deluxetable}{ccccccc}
\tablecaption{$Y$(SMC)
\label{YSMC}}
\tablewidth{0pt}
\tablehead{\colhead{$T_e$(K)} & \ & \multicolumn{5}{c}{$N_e$(cm$^{-3}$)} \\
&& \colhead{53}  & \colhead{100} & \colhead{143}
 & \colhead{162} & \colhead{247}}
\startdata
11200 && 0.2431 & 0.2416 & 0.2404 & 0.2399 & 0.2377\tablenotemark{a} \\
11800 && 0.2435 & 0.2419 & 0.2405 & 0.2399\tablenotemark{a} & 0.2375 \\
11950 && 0.2436 & 0.2420 & 0.2405\tablenotemark{a} & 0.2399 & 0.2374 \\
12400 && 0.2439 & 0.2421\tablenotemark{a} & 0.2406 & 0.2399 & 0.2372 \\
13000 && 0.2443\tablenotemark{a} & 0.2423 & 0.2407 & 0.2400 & 0.2370 \\
\enddata
\tablenotetext{a}{Entries that correspond to minimum $\chi^2$ values, see
Table~\ref{chi}.} 
\end{deluxetable}

\clearpage

\begin{deluxetable}{ccr@{$\pm$}lr@{$\pm$}l}
\tablecaption{Chemical Composition\tablenotemark{a}
\label{Composition}}
\tablewidth{0pt}
\tablehead{\colhead{Element} &
    \multicolumn{3}{c}{NGC~346\tablenotemark{b}} &
        \multicolumn{2}{c}{M17\tablenotemark{b,c}} \\
& \colhead{$t^2=0.0013$} & \multicolumn{2}{c}{$t^2=0.0220$}}
\startdata
$Y$                & 0.2445~ & 0.24050 
                                       & 0.00180 & 0.2797 & 0.0060 \\
$Z$                & 0.00263 & 0.00315 & 0.00063 & 0.0212 & 0.0030 \\
$O$                & 0.00142 & 0.00171 & 0.00025 & 0.0083 & 0.0012 \\
\\
$\Delta Y/\Delta Z$ &   ...   &   1.9   &   0.5   &  2.13  &  0.5   \\
$\Delta Y/\Delta O$ &   ...   &   3.5   &   0.9   &  5.45  &  1.1   \\
\enddata
\tablenotetext{a}{Given by mass.}
\tablenotetext{b}{This paper.}
\tablenotetext{c}{\citealt{pei92,est99}.}
\end{deluxetable}

\clearpage

\begin{deluxetable}{ccccccc}
\tablecaption{$Y_p$ Derived from the SMC
\label{YpSMC}}
\tablewidth{0pt}
\tablehead{\colhead{$T_e$(K)} & \ & \multicolumn{5}{c}{$N_e$(cm$^{-3}$)} \\
&& \colhead{53}  & \colhead{100} & \colhead{143}
 & \colhead{162} & \colhead{247}}
\startdata
11200 && 0.2363 & 0.2348 & 0.2336 & 0.2331 & 0.2309\tablenotemark{a} \\
11800 && 0.2373 & 0.2357 & 0.2343 & 0.2337\tablenotemark{a} & 0.2313 \\
11950 && 0.2376 & 0.2360 & 0.2345\tablenotemark{a} & 0.2339 & 0.2314 \\
12400 && 0.2384 & 0.2366\tablenotemark{a} & 0.2351 & 0.2344 & 0.2317 \\
13000 && 0.2395\tablenotemark{a} & 0.2375 & 0.2359 & 0.2352 & 0.2322 \\
\enddata
\tablenotetext{a}{Entries that correspond to minimum $\chi^2$ values, see
Table~\ref{chi}.} 
\end{deluxetable}

\clearpage

\begin{figure}
\includegraphics{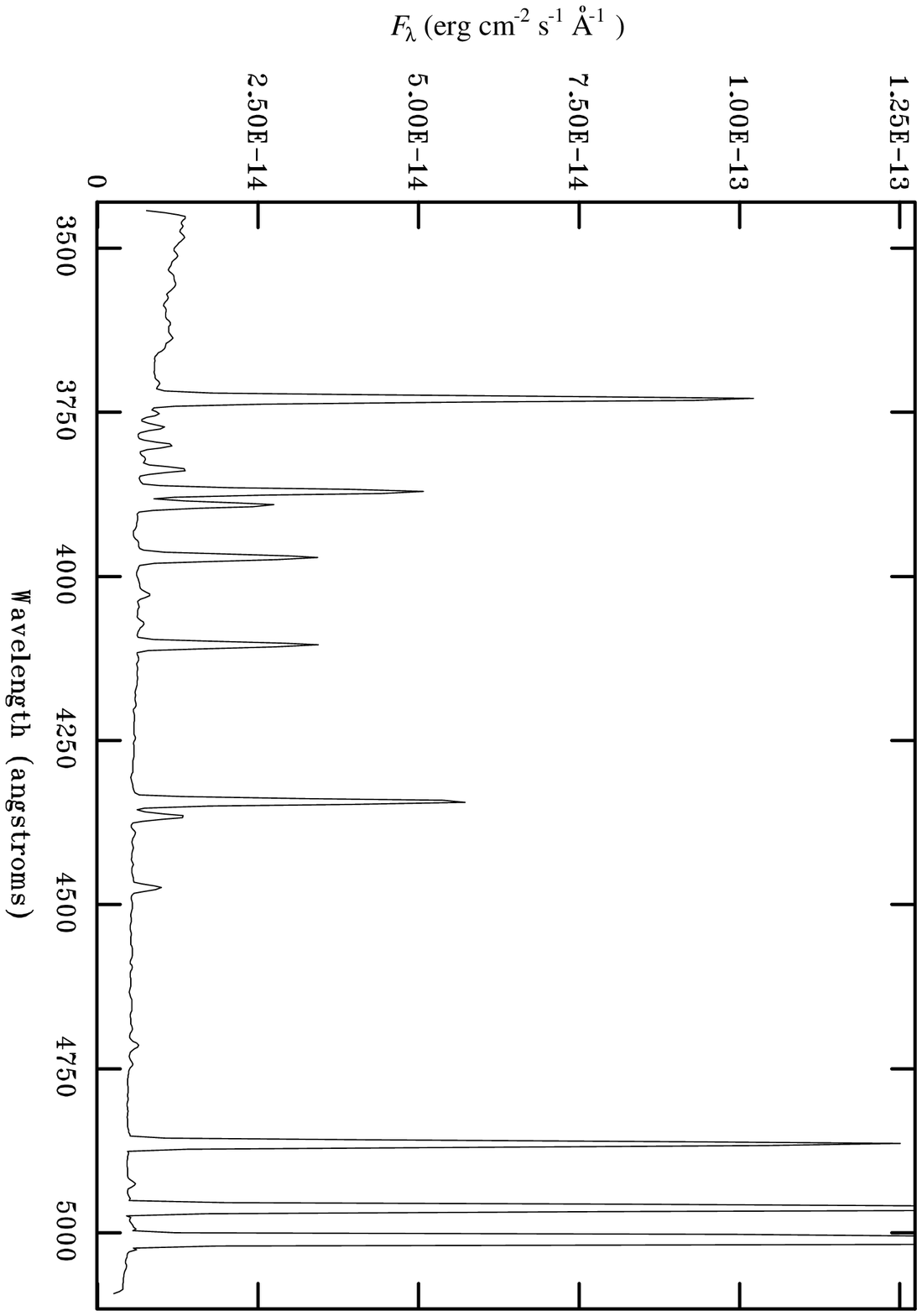}
\end{figure}

\figcaption[f1.eps]{
\label{blue}
Blue spectrum for region A.
}

\begin{figure}
\includegraphics{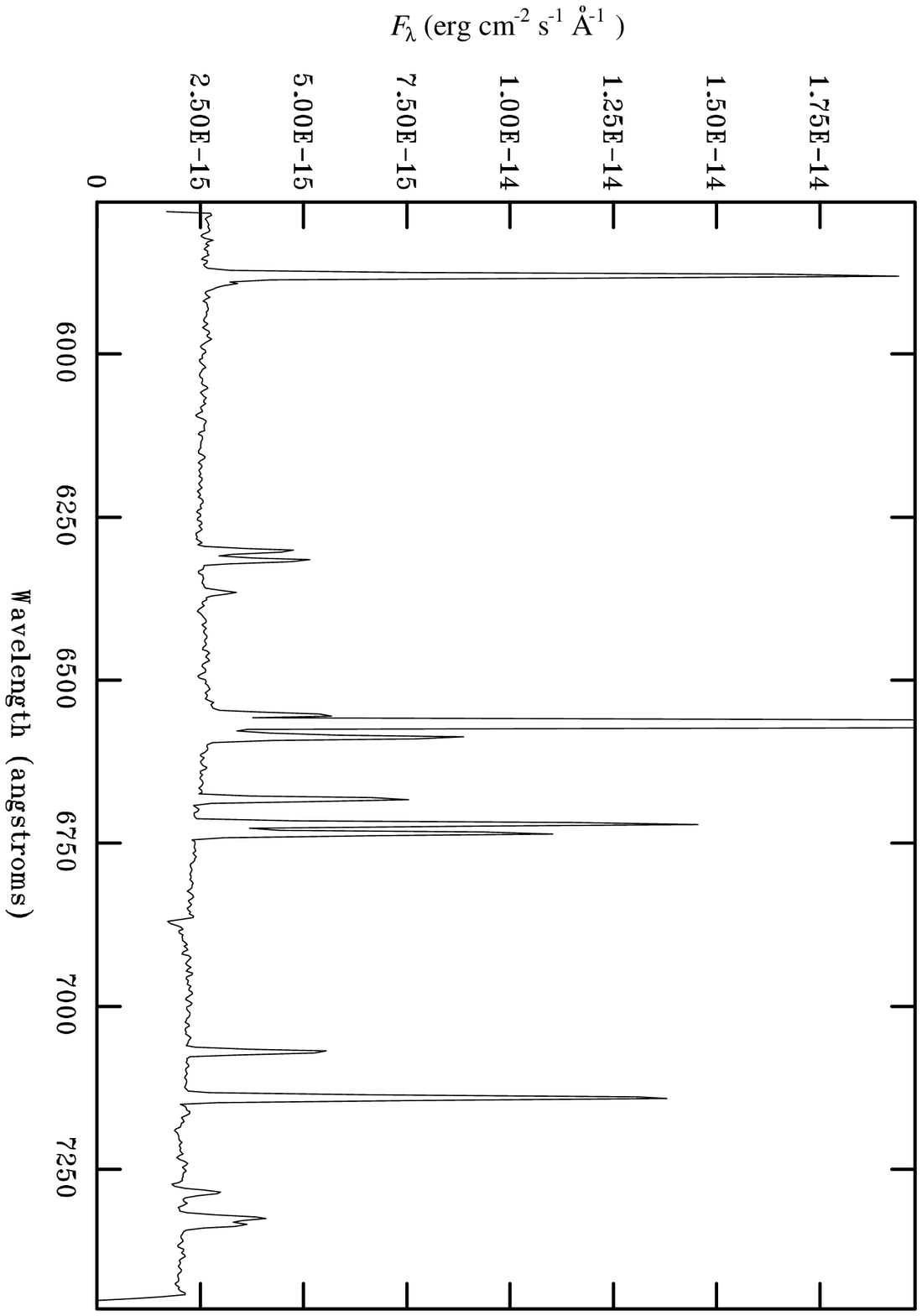}
\end{figure}

\figcaption[f2.eps]{
\label{red}
Red spectrum for region A.
}

\begin{figure}
\includegraphics{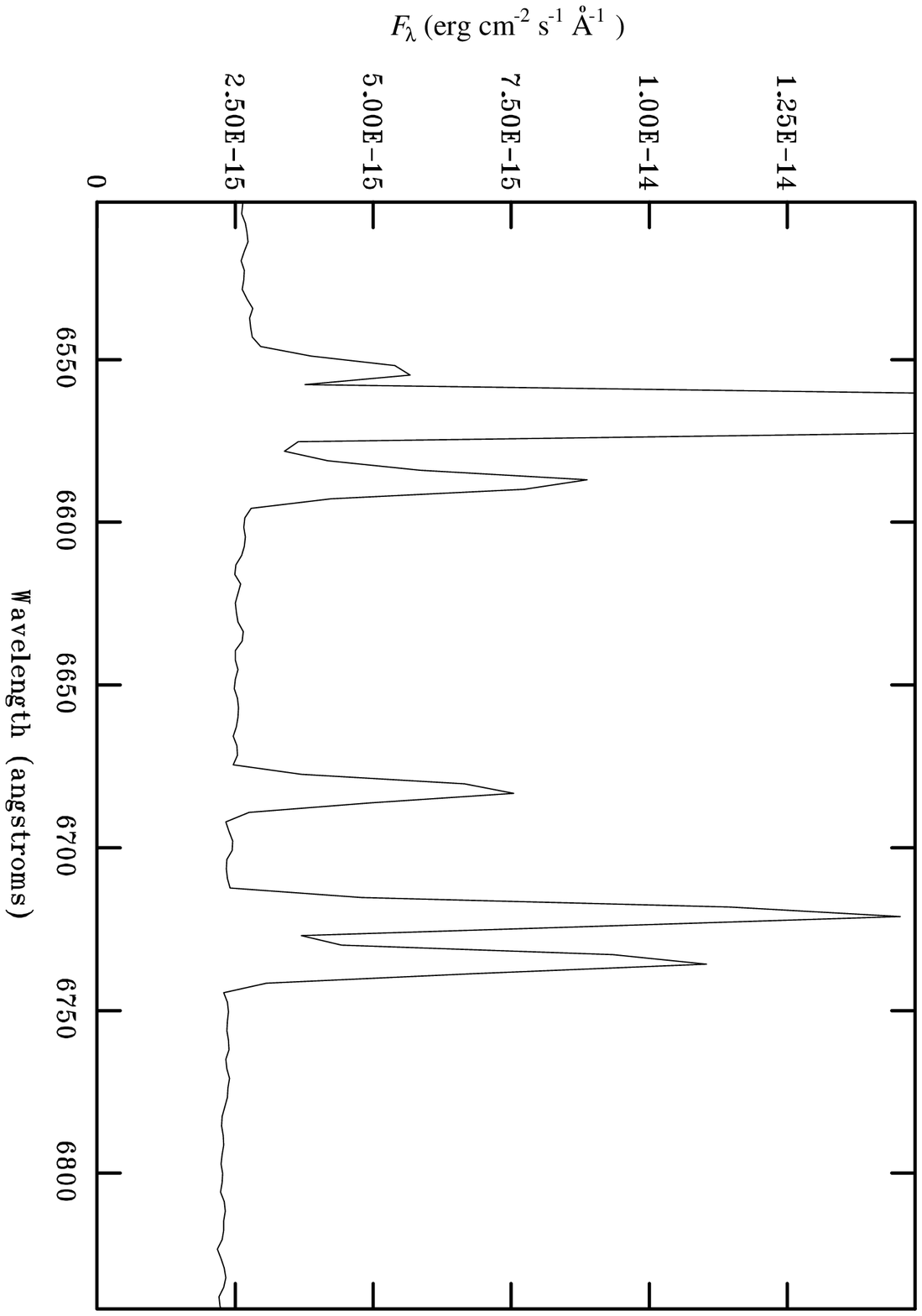}
\end{figure}

\figcaption[f3.eps]{
\label{red-detail}
Spectrum of region A near H$\alpha$ that shows [N~{\sc{ii}}], [S~{\sc{ii}}],
and He~{\sc{i}} lines.
}

\begin{figure}
\includegraphics{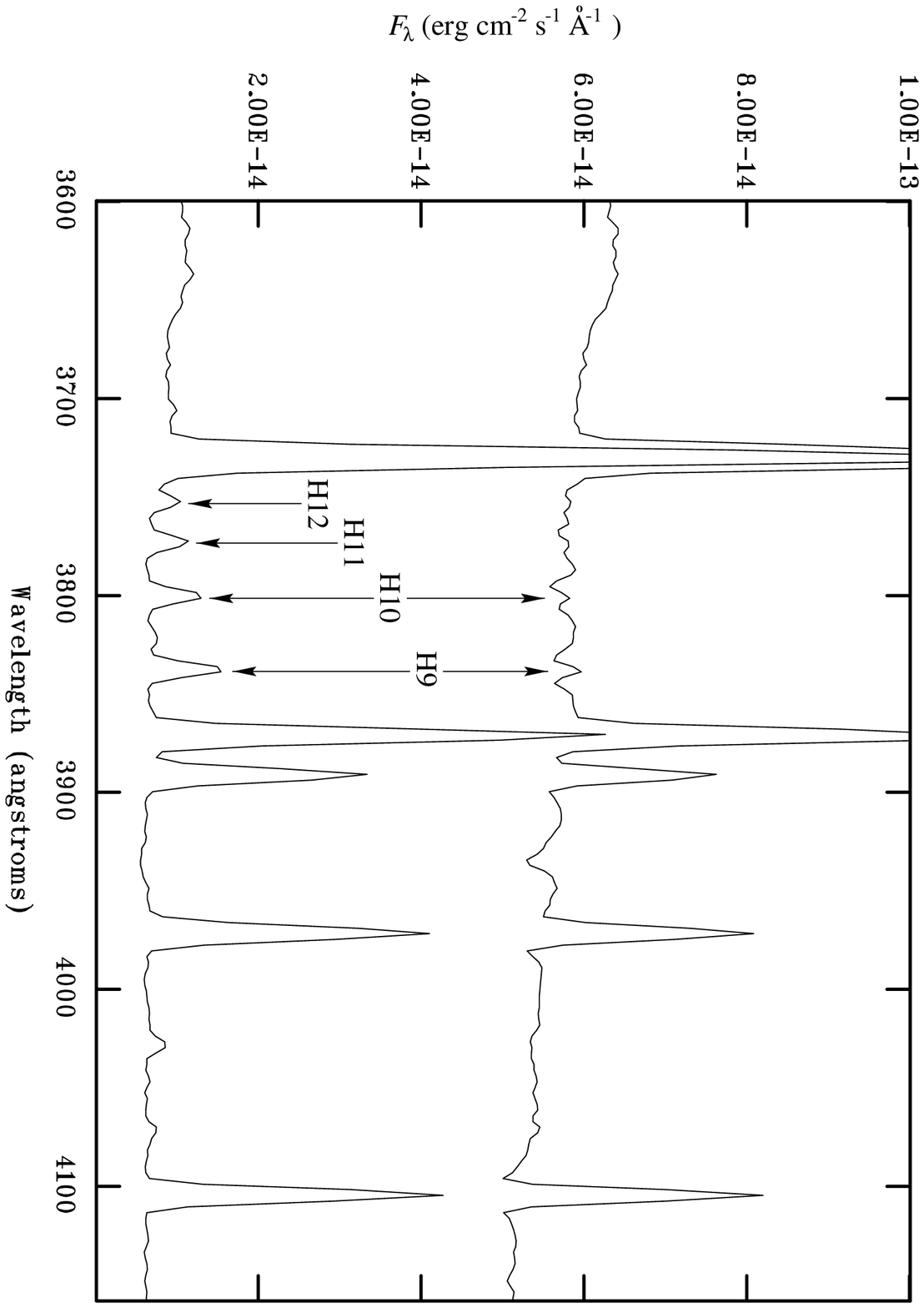}
\end{figure}

\figcaption[f4.eps]{
\label{blue-detail}
Spectra of NGC 346 with and without underlying absorption. The vertical scale
is for the lower spectrum (region~A). The flux of the upper spectrum
(region~B) was normalized to the H$\alpha$ emmision line flux for the lower
spectrum.
}

\begin{figure}
\scalebox{.875}{
\includegraphics{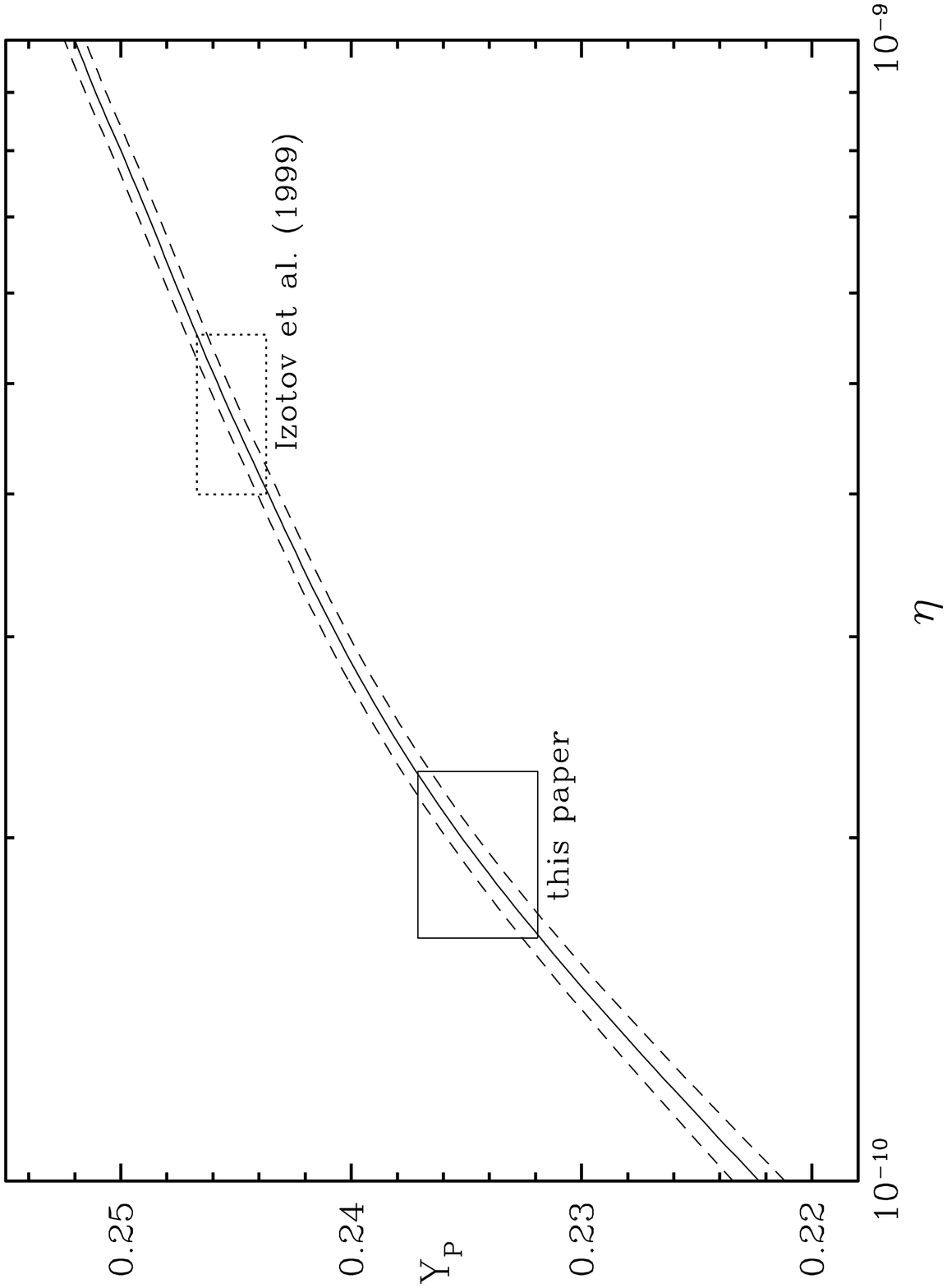}
}
\end{figure}

\figcaption[f5.eps]{
\label{eta}
$Y_p$ versus $\eta$ diagram from the Big Bang nucleosynthesis computations by
\citet{cop95}. The low $Y_p$ box is by us while the high $Y_p$ box is by
\citet{izo99}. The dashed curves correspond to $1 \sigma$ error bars, the boxes
include $1 \sigma$ errors in the observations and the theoretical cmputations
added in quadrature.}

\end{document}